\documentclass[paper,toc,nohyper]{JHEP3}
\usepackage{epsfig,cite}
\usepackage{amsbsy} 
\usepackage{graphicx}
\graphicspath{{plots/}} % TeX will look for pix here

\newcommand\new{\newcommand}         % shorthand for \newcommand
\def\beq{\begin{equation}}   
\def\eeq{\end{equation}}
\def\bea{\begin{eqnarray}}  
\def\eea{\end{eqnarray}}

\def\O{y}

\def\CA{C_A}

\def\NF{N_F}

\def\bt{B_T}
\def\bw{B_W}

\def\d{\hbox{d}}

\def\ln{\hbox{ln}}

\renewcommand{\textfraction}{0.0}

\new{\emem}{{\ifmmode\mathrm{e}^-\else e$^-$\fi}}
\new{\epem}{{\ifmmode\mathrm{e}^+\else e$^+$\fi}}
\new{\zo}  {{\ifmmode\mathrm{Z}\else Z\fi}}
\new{\epm} {{\ifmmode\mathrm{e^+e^-}\else $\mathrm{e^+e^-}$\fi}}
\new{\qq}  {{\ifmmode\mathrm{q}\else q\fi}}
\new{\qqb} {{\ifmmode\bar{\mathrm{q}}\else $\bar{\mathrm{q}}$\fi}}
\new{\bq}  {{\ifmmode\mathrm{b}\else b\fi}}
\new{\bqb} {{\ifmmode\bar{\mathrm{b}}\else $\bar{\mathrm{b}}$\fi}}
\new{\qqbar}{\qq\qqb}

\new{\LEP}        {\mbox{\small\textsc{LEP}}}
\new{\LEPONE}     {\mbox{\small\textsc{LEP1}}}
\new{\LEPTWO}     {\mbox{\small\textsc{LEP2}}}
\new{\CERN}       {\mbox{\small\textsc{CERN}}}
\new{\ALEPH}      {\mbox{\small\textsc{ALEPH}}}
\new{\DELPHI}     {\mbox{\small\textsc{DELPHI}}}
\new{\LD}         {\mbox{\small\textsc{L3}}}
\new{\OPAL}       {\mbox{\small\textsc{OPAL}}}

\new{\eV}         {{\ifmmode {\mathrm{ eV}}\else ${\mathrm{ eV}}$\fi}}
\new{\MeV}        {{\ifmmode {\mathrm{ MeV}}\else ${\mathrm{ MeV}}$\fi}}
\new{\MeVc}       {{\ifmmode {\mathrm{ MeV}}/c\else ${\mathrm{ MeV}}/c$\fi}}
\new{\MeVcc}      {{\ifmmode {\mathrm{ MeV}}/c^2\else ${\mathrm{ MeV}}/c^2$\fi}}
\new{\GeV}        {{\ifmmode {\mathrm{ GeV}}\else ${\mathrm{ GeV}}$\fi}}
\new{\GeVc}       {{\ifmmode {\mathrm{ GeV}}/c\else ${\mathrm{GeV}}/c$\fi}}
\new{\GeVcc}      {{\ifmmode {\mathrm{ GeV}}/c^2\else ${\mathrm{GeV}}/c^2$\fi}}
\new{\TeV}        {{\ifmmode {\mathrm{ TeV}}\else ${\mathrm{ TeV}}$\fi}}

\new{\JS}         {\mbox{\small\textsc{JETSET}}}
\new{\HW}         {\mbox{\small\textsc{HERWIG}}}
\new{\AR}         {\mbox{\small\textsc{ARIADNE}}}
\new{\PY}         {\mbox{\small\textsc{PYTHIA}}}
\new{\JSv}        {\mbox{\small\textsc{JETSET\ 7.405}}}
\new{\HWo}        {\mbox{\small\textsc{HERWIG\ 5.8}}}
\new{\HWn}        {\mbox{\small\textsc{HERWIG\ 5.9}}}
\new{\ARv}        {\mbox{\small\textsc{ARIADNE\ 4.05}}}
\new{\PYv}        {\mbox{\small\textsc{PYTHIA\ 5.7}}}

\new{\Mh}         {{\ifmmode M_{\mathrm{ H}}
                    \else $M_{\mathrm{H}}$\fi}}

\new{\Mz}         {{\ifmmode M_{\mathrm{Z}}
                    \else $M_{\mathrm{Z}}$\fi}}
\new{\Mzsq}       {{\ifmmode M^2_{\mathrm{ Z}}
                    \else $M^2_{\mathrm{Z}}$\fi}}

\new{\as}[1]      {{\ifmmode\alpha^{#1}_s
                    \else$\alpha^{#1}_s$\fi}}
\new{\asx}[1]      {{\ifmmode a^{#1}_s
                    \else $a^{#1}_s$\fi}}
\new{\asb}[1]     {{\ifmmode\overline{\alpha}^{#1}_s
                    \else $\overline{\alpha}^{#1}_s$\fi}}
\new{\asmz}       {{\ifmmode\alpha_s(\Mzsq)
                    \else $\alpha_s(\Mzsq)$\fi}}
\new{\lqcd}       {{\ifmmode\Lambda_{\mathrm{ QCD}}
                    \else $\Lambda_{\mathrm{ QCD}}$\fi}}
\new{\lqcdsq}     {{\ifmmode\Lambda^2_{\mathrm{ QCD}}
                    \else $\Lambda^2_{\mathrm{ QCD}}$\fi}}
\new{\llla}       {{\ifmmode\Lambda_{\mathrm{ LLA}}
                    \else $\Lambda_{\mathrm{ LLA}}$\fi}} 
\new{\lmsbar}[1]  {{\ifmmode \Lambda^{(#1)}_{\overline{\mathrm{MS}}}
                    \else $\Lambda^{(#1)}_{\overline{\mathrm{MS}}}$\fi}}
\new{\lmsb}       {{\ifmmode \Lambda_{\overline{\mathrm{MS}}}
                    \else $\Lambda_{\overline{\mathrm{MS}}}$\fi}}
\new{\lmsbsq}     {{\ifmmode \Lambda^{2}_{\overline{\mathrm{MS}}}
                    \else $\Lambda^{2}_{\overline{\mathrm{MS}}}$\fi}}

\title{\boldmath 
First determination of the strong coupling constant using 
NNLO predictions for hadronic event shapes in  $\mathrm{e}^+\mathrm{e}^-$ annihilations
}
\author{G.\ Dissertori\\
Institute for Particle Physics, ETH Zurich,\\
      8093 Zurich, Switzerland\\
	E-mail: \email{dissertori@phys.ethz.ch}}
\author{
A.~Gehrmann--De Ridder\\
Institute for Theoretical Physics, ETH Zurich,\\
      8093 Zurich, Switzerland\\
E-mail: \email{gehra@phys.ethz.ch}}
\author{
T.~Gehrmann\\
Institut f\"ur Theoretische Physik, Universit\"at Z\"urich,
Winterthurerstrasse 190,\\ CH-8057 Z\"urich, Switzerland\\
E-mail: \email{thomas.gehrmann@physik.unizh.ch}}
\author{E.W.N.~Glover\\
Institute of Particle Physics Phenomenology, 
        Department of Physics,\\
        University of Durham, Durham, DH1 3LE, UK\\
	E-mail: \email{e.w.n.glover@durham.ac.uk}}
\author{
G.~Heinrich\\
School of Physics, The University of Edinburgh, Edinburgh EH9 3JZ,
UK\\
E-mail: \email{gheinric@ph.ed.ac.uk}}
\author{
H.~Stenzel\\
II. Physikalisches Institut, Justus-Liebig Universit\"at Giessen\\
Heinrich-Buff Ring 16, D-35392 Giessen, Germany\\
E-mail: \email{Hasko.Stenzel@exp2.physik.uni-giessen.de}}

\abstract{
We present the first determination of the strong coupling constant from a fit of
next-to-next-to-leading order QCD predictions to event-shape variables, measured 
in \epm\ annihilations at \LEP. The data have been collected
by the \ALEPH\ detector at centre-of-mass energies between 91 and 206 \GeV. Compared
to results of next-to-leading order fits we observe that the central fit values are
lower by about $10\,\%$, with considerably reduced scatter among the results obtained 
with different
event-shape variables. The dominant systematic uncertainty from renormalization scale
variations is reduced by a factor of two. By combining the results for several event-shape variables
and centre-of-mass energies, we find
\begin{center} 
    $\asmz = 0.1240 \;\pm\; 0.0008\,\mathrm{(stat)}
     					 \;\pm\; 0.0010\,\mathrm{(exp)}
                                   \;\pm\; 0.0011\,\mathrm{(had)}
                                   \;\pm\; 0.0029\,\mathrm{(theo)} $.
\end{center}
}

\preprint{{ZU-TH 28/07}, IPPP/07/91, 
{ETHZ-IPP RP-2007-04}, Edinburgh 2007-48}

\begin{document}

%%%%%%%%%%%%%%%%%%%%%%%%%%%%%%%%%%%%%%%%%%%%%%%%%
%%%%% Introduction
%%%%%%%%%%%%%%%%%%%%%%%%%%%%%%%%%%%%%%%%%%%%%%%%%%

\section{Introduction}
\label{sec:intro}

Quantum Chromodynamics (QCD) is generally accepted to be the correct theory for the
description of strong interactions between quarks and gluons \cite{QCDbooks}. If the 
quark masses and mixing angles 
are fixed, then the only free parameter of the theory is the strong coupling constant, \as{}. 
Therefore it is of paramount importance to measure this parameter at the best possible precision.
In particular, measurements based on different underlying processes, at different energy scales, constitute
an important consistency check of the theory and are used to prove the so-called running of the
coupling constant, ie., the decrease of the coupling strength with increasing energy scale \cite{Gross-Wilczek-Politzer}.
During the last twenty years an enormous wealth of measurements has become available, using
many different processes with different initial and final states~\cite{as-overview,pdg}. These measurements resulted
in values for $\as{}(Q^2)$, at different scales $Q$, which are perfectly consistent with the expected running of the
coupling. When evolved to a single scale, most commonly chosen to be the mass of the Z boson, \Mz, the 
measurements agree with each other within a few percent, and thus can be used to extract a
world-average value. The particle data group~\cite{pdg} quotes 
$\asmz=0.1176\pm 0.0020$, which has been improved upon in a more recent 
global analysis to $\asmz=0.1189 \pm 0.0010$ \cite{as-worldaverage}.

Several classes of observables can be identified as useful for an \as{} measurement. \textit{Inclusive observables},
such as the ratio of the cross sections for $\epm\rightarrow\,\mathrm{hadrons}$ and $\epm\rightarrow\mu^+\mu^-$
or sum rules in deep-inelastic scattering, do not resolve the structure of the hadronic final state. The theoretical
predictions are obtained with perturbative QCD (pQCD) and typically known up to next-to-next-to-leading order (NNLO) 
in the strong coupling constant. Non-perturbative effects related to the transition (hadronisation) of partons (quarks and gluons) to hadrons are strongly suppressed by the scale $Q$ of the process. \textit{Exclusive observables}, such as 
event-shape variables in \epm\ annihilations, discussed in detail below, resolve topological properties of the
hadronic final state and thus are sensitive to gluon radiation. This class of observables shows stronger sensitivity
to hadronisation effects, at least in phase-space regions characterised by soft and collinear gluon radiation.
Finally, \textit{spectroscopic properties} of mesons and baryons, such as masses or energy level splittings of heavy quark-antiquark
resonances, are necessarily determined by non-perturbative phenomena. Major recent progress in calculations within non-relativistic QCD and 
 lattice QCD  
has resulted in very precise \as{}\  determinations from this last type of observables \cite{lattice}.

Thanks to the clean initial state and the very large event statistics, \epm\ 
annihilations to hadrons at the LEP collider
at centre-of-mass energies between 91 and 206 \GeV\ constituted an excellent laboratory 
for precision tests of QCD in
general, and precise measurements of the strong coupling constant in 
particular~\cite{aleph,opal,l3,delphi}.
 These are complemented by data obtained 
by the SLD experiment~\cite{sld} 
at SLAC  at centre-of-mass energy of  91 \GeV\ 
and by data obtained at the  DESY PETRA collider at lower energies, 
especially the reanalysis of data from the JADE experiment~\cite{jade}. 

Many of those measurements were
based on event-shape variables such as thrust \cite{thrust}. These variables probe the structure of the hadronic
final state. In the leading-order picture, \epm\ annihilation to hadrons occurs via $\epm\rightarrow\qqbar$ and subsequent
hadronisation to stable hadrons, resulting in a back-to-back two-jet structure of the event. At the next order of
perturbation theory, gluon radiation off quarks will lead to deviations from this two-jet structure due to the appearance
of additional jets. In general, event-shape variables quantify the structure of an event by a single measure. 

Until recently, pQCD predictions for event-shape observables were known up to next-to-leading order
(NLO) in the strong coupling. As a consequence, the residual sensitivity to the choice of the unphysical renormalisation
scale, $\mu$, resulted in systematic uncertainties of the \as{}\ measurements at the 10\% level. Those systematic
uncertainties are typically estimated by varying $\mu$ over a broad range, eg., 
$1/2 < \mu/Q < 2$, since this
variation probes most of the missing higher-order contributions. Using a very large set of event shapes, the
\DELPHI\ collaboration observed a large spread in the \asmz\ values extracted from the different observables
\cite{Delphi-as-spread}, when 
fixing $\mu = \Mz$. In addition, the measurements preferred rather large values of $\asmz \approx 0.125 - 0.13$ 
compared to the world average. 
On the other hand, simultaneously fitting \asmz\ and the scale $\mu$ resulted in a considerably
reduced spread of \asmz\ values, 
however, at the price of a large spread of preferred $\mu$ parameters, which in addition
turned out to be rather small compared to the ``natural'' choice $\mu = \Mz$.  Similar observations were made by the
other \LEP\ collaborations and on the  SLD data~\cite{burrows}. 

An important part of the missing
higher-order contributions was identified to be due to large logarithms in the 
event-shape variable, related to
the incomplete cancellation of real and virtual corrections due to gluon radiation 
\cite{NLLA-reference-1, NLLA-reference-2, NLLA-reference-3, NLLA-reference-4, NLLA-reference-5, NLLA-reference-6}. 
Once  these logarithms are resummed to all orders in pQCD and consistently combined
with the fixed-order calculations, the resulting improved predictions lead to a better description of the event-shape
variables, most notably close to the two-jet region of the phase space. 
Consequently, the \as{}\ measurements
from fits of these improved predictions were characterised by reduced systematic 
uncertainties (now at the 5\% level)
and showed better agreement with other determinations, with typical values in the range $\asmz=0.118 - 0.125$.
Other systematic uncertainties, from experimental effects and most notably from the estimation of non-perturbative contributions, were smaller than those from the $\mu$-scale variations. Non-perturbative hadronisation corrections are 
estimated by either using phenomenological models (string or cluster fragmentation), implemented in Monte
Carlo simulations \cite{Pythia, Herwig, Ariadne}, or by adding power-law corrections to the 
purely perturbative predictions \cite{power-law}.

From all this it became clear that major progress in the pQCD calculations was 
necessary in order to 
push the uncertainties of \as{}\ measurements from event shapes below 
the 5\% range, approaching 
the precision obtained with inclusive observables or lattice calculations. 
Such major progress
has recently been made with the calculation of the NNLO corrections to event-shape 
variables \cite{ourt,ourevent}.
Consequently, it is of great interest to include these new corrections in fits for \asmz. In this paper we
describe first fits of this kind and discuss the results. The measurements are 
obtained with event-shape distributions
measured by the \ALEPH\ experiment \cite{ALEPH-qcdpaper} at centre-of-mass energies
between 91 and 206 \GeV. We concentrate on the comparison of the results found at 
NNLO with those at NLO and NLO matched to resummation of leading and next-to-leading 
logarithms (NLLA), as used in~\cite{ALEPH-qcdpaper}.
In a forthcoming publication results using NNLO matched to NLLA 
will be discussed in detail.

%%%%%%%%%%%%%%%%%%%%%%%%%%%%%%%%%%%%%%%%%%%%%%%%%
%%%%% theory
%%%%%%%%%%%%%%%%%%%%%%%%%%%%%%%%%%%%%%%%%%%%%%%%%%

\section{Theoretical predictions}
\label{sec:theory}

The perturbative expansion for the distribution of a 
generic observable $\O$ up to NNLO at \epm\ centre-of-mass energy $\sqrt{s}$, 
for a renormalisation scale $\mu^2 = s$ and 
$\alpha_s\equiv \alpha_s(s)$  is given by
\begin{eqnarray}
\frac{1}{\sigma_{{\rm had}}}\, \frac{\d\sigma}{\d \O} &=& 
\left(\frac{\alpha_s}{2\pi}\right) \frac{\d \bar A}{\d \O} +
\left(\frac{\alpha_s}{2\pi}\right)^2 \frac{\d \bar B}{\d \O}
+ \left(\frac{\alpha_s}{2\pi}\right)^3 
\frac{\d \bar C}{\d \O} + {\cal O}(\alpha_s^4)\;.
\label{eq:NNLO}
\end{eqnarray}
Here the event-shape distribution  
is normalised to the total hadronic cross section $\sigma_{\rm{had}}$.
With the assumption of massless quarks we have at NNLO
  \begin{equation}
  \sigma_{\rm{had}}=\sigma_0\,
\left(1+\frac{3}{2}C_F\,\left(\frac{\alpha_s}{2\pi}\right)
+K_2\,\left(\frac{\alpha_s}{2\pi}\right)^2+{\cal O}(\alpha_s^3)\,
\right) \;,
\end{equation}
where the Born cross section for  $\epm\rightarrow\qqbar$ (photon 
exchange only, with electromagnetic coupling $\alpha$ and quark charge 
$e_q$) is
\begin{equation}
\sigma_0 = \frac{4 \pi \alpha}{3 s} N \, e_q^2\;.
\end{equation}
The constant $K_2$ is given by,
\begin{equation}
  K_2=\frac{1}{4}\left[- \frac{3}{2}C_F^2
+C_FC_A\,\left(\frac{123}{2}-44\zeta_3\right)+C_FT_RN_F\,(-22+16\zeta_3)
 \right] \;,
\end{equation}
where $\zeta_3=1.202056903\ldots$, and the QCD colour factors are
\begin{equation}
\CA = N,\qquad C_F = \frac{N^2-1}{2N},
\qquad T_R = \frac{1}{2}\; 
\end{equation}
for $N=3$ colours and $N_F$ light quark flavours.

In practice, we use the perturbative coefficients $A$, $B$ and $C$, 
computed in the $\overline{{\rm MS}}$-scheme in~\cite{ourevent}, which are 
all normalised to 
$\sigma_0$:
\begin{eqnarray}
\frac{1}{\sigma_0}\, \frac{\d\sigma}{d \O} &=& 
\left(\frac{\alpha_s}{2\pi}\right) \frac{\d  A}{\d \O} +
\left(\frac{\alpha_s}{2\pi}\right)^2 \frac{\d  B}{\d \O}
+ \left(\frac{\alpha_s}{2\pi}\right)^3 
\frac{\d  C}{\d \O} + {\cal O}(\alpha_s^4)\,.
\label{eq:NNLOsigma0}
\end{eqnarray}
However, $A$, $B$ and $C$ are straightforwardly related to $\bar{A}$, $\bar{B}$ 
and $\bar{C}$,
\begin{eqnarray}
\bar{A} &=& A\;,\nonumber \\
\bar{B} &=& B - \frac{3}{2}C_F\,A\;,\nonumber \\
\bar{C} &=& C -  \frac{3}{2}C_F\,B+ \left(\frac{9}{4}C_F^2\,-K_2\right)\,A 
\;.\label{eq:ceff}
\end{eqnarray} 
These coefficients are computed at a renormalisation scale fixed to 
the centre-of-mass energy, 
  and 
depend therefore only on the value of the observable $y$.
They explicitly include only QCD corrections with non-singlet 
quark couplings and are therefore independent of electroweak 
couplings.
At ${\cal O}(\alpha_s^2)$, these amount to the full 
corrections, while the ${\cal O}(\alpha_s^3)$ corrections also 
receive a pure-singlet contribution. This pure-singlet contribution 
arises from the interference of diagrams where the external gauge boson 
couples to different quark lines. In four-jet observables at 
 ${\cal O}(\alpha_s^3)$, these singlet contributions were found to be 
extremely small~\cite{dixonsigner}.

The QCD coupling constant evolves according to the renormalisation group 
equation, which is to NNLO:
\begin{equation}
\mu^2 \frac{\d \alpha_s(\mu^2)}{\d \mu^2} = -\alpha_s(\mu^2) 
\left[\beta_0 \left(\frac{\alpha_s(\mu^2)}{2\pi}\right) 
+ \beta_1 \left(\frac{\alpha_s(\mu^2)}{2\pi}\right)^2 
+ \beta_2 \left(\frac{\alpha_s(\mu^2)}{2\pi}\right)^3 
+ {\cal O}(\alpha_s^4) \right]\,
\end{equation}
with the $\overline{{\rm MS}}$-scheme coefficients
\begin{eqnarray}
\beta_0 &=& \frac{11\, \CA - 4 \, T_R \NF}{6}\;,\nonumber  \\
\beta_1 &=& \frac{17\, \CA^2 - 10\, C_A T_R \NF- 6\, C_F T_R \NF}{6}\;, \nonumber \\
\beta_2 &=&\frac{1}{432}
\big( 2857\, C_A^3 + 108\, C_F^2 T_R N_F -1230\, C_FC_A T_R N_F
-2830\, C_A^2T_RN_F \nonumber \\ &&
+ 264\, C_FT_R^2 N_F^2 + 316 \, C_AT_R^2N_F^2\big)\;.
\end{eqnarray}

The above equation
is solved by introducing $\Lambda_{\overline{{\rm MS}}}^{(N_F)}$ 
as integration constant
with $L= 2\, \log(\mu/\Lambda_{\overline{{\rm MS}}}^{(N_F)})$, 
yielding the running coupling constant:
\begin{equation}
\alpha_s(\mu^2) = \frac{2\pi}{\beta_0 L}\left( 1- 
\frac{\beta_1}{\beta_0^2}\, \frac{\log L}{L} + \frac{1}{\beta_0^2 L^2}\,
\left( \frac{\beta_1^2}{\beta_0^2}\left( \log^2 L - \log L - 1
\right) + \frac{\beta_2}{\beta_0}  \right) \right)\;.
\label{eq:runningas}
\end{equation}

In terms of the running coupling $\as{}(\mu^2)$, the 
NNLO expression for event-shape distributions becomes
\begin{eqnarray}
\frac{1}{\sigma_{{\rm had}}}\, \frac{\d\sigma}{\d y} (s,\mu^2,y) &=& 
\left(\frac{\as{}(\mu^2)}{2\pi}\right) \frac{\d \bar A}{\d y} +
\left(\frac{\as{}(\mu^2)}{2\pi}\right)^2 \left( 
\frac{\d \bar B}{\d y} + \frac{\d \bar A}{\d y} \beta_0 
\log\frac{\mu^2}{s} \right)
\nonumber \\ &&
+ \left(\frac{\as{}(\mu^2)}{2\pi}\right)^3 
\bigg(\frac{\d \bar C}{\d y} + 2 \frac{\d \bar B}{\d y}
 \beta_0\log\frac{\mu^2}{s}
\nonumber \\ &&
\hspace{24mm} + \frac{\d \bar A}{\d y} \left( \beta_0^2\,\log^2\frac{\mu^2}{s}
+ \beta_1\, \log\frac{\mu^2}{s}   \right)\bigg)\nonumber \\
&&+ {\cal O}(\as{4})  \;.
\label{eq:NNLOmu} 
\end{eqnarray}

The coefficients $A, B$ and $C$ have been computed for several event-shape
variables~\cite{ourt,ourevent}. The calculation is carried out using 
a newly developed
parton-level event generator programme {\tt EERAD3} which contains 
the relevant 
matrix elements with up to five external partons~\cite{3jme,muw2,V4p,tree5p}. 
Besides explicit infrared divergences from the loop integrals, the 
four-parton and five-parton contributions yield infrared divergent 
contributions if one or two of the final state partons become collinear or 
soft. In order to extract these infrared divergences and combine them with 
the virtual corrections, the antenna subtraction method~\cite{ant} 
was extended to NNLO level~\cite{ourant} and implemented
for $\epm \to 3\,\mathrm{jets}$ and related event-shape variables~\cite{eerad3}. The analytical cancellation of all 
infrared divergences serves as a very strong check on the implementation. 
{\tt EERAD3} yields the perturbative  $A$, $B$ and $C$ coefficients as 
histograms for all infrared-safe event-shape variables related to three-particle 
final states at leading order. 
As a cross check, the $A$ and $B$  coefficients have also been obtained from an independent integration~\cite{event2}
of the NLO matrix elements~\cite{ERT}, showing excellent agreement. While $A$ and $B$ 
can be computed to very high numerical accuracy, the computation of  
$C$ is so CPU-intensive that we could obtain only results with numerical integration errors of 
typically a few percent. These numerical
 errors on the $C$-coefficient will be included in the 
analysis of $\alpha_s$ as described below. 

For small values of $y$, the fixed-order expansion, eq.\ (\ref{eq:NNLOmu}), fails to converge, 
because the fixed-order coefficients are enhanced by powers of $\ln(1/y)$,
$y A(y) \sim \ln y$, 
$y B(y) \sim \ln^3 y$ and $y C(y) \sim \ln^5 y$. 
In order to obtain reliable predictions
in the region of $y \ll 1$ it is necessary to resum entire sets of logarithmic terms at all orders in \as{}. 
A detailed description of the predictions at next-to-leading-logarithmic approximation (NLLA) can
be found in Ref.\ \cite{as_theory-uncertainties}. There the matching of the NLO and NLLA
expressions, which is necessary in order to avoid double counting, is also described. The extension of 
this matching procedure to NNLO with NNLA is
outlined in~\cite{NLLA-reference-2} 
 and will be discussed in detail in a 
separate publication.

%%%%%%%%%%%%%%%%%%%%%%%%%%%%%%%%%%%%%%%%%%%%%%%%%
%%%%% Data sets and observables
%%%%%%%%%%%%%%%%%%%%%%%%%%%%%%%%%%%%%%%%%%%%%%%%%%

\section{Observables and data sets}
\label{sec:data}
\renewcommand{\arraystretch}{1.1}
\renewcommand{\textfraction}{0.0001}

We have studied the six event-shape distributions thrust $T$ \cite{thrust}, heavy jet mass $M_H$ \cite{heavyjetmass},
total and wide jet broadening ($\bt, \bw$) \cite{broadenings}, C-parameter $C$ \cite{cparameter}
and the Durham jet resolution parameter $y_3$ \cite{Durhamythree}. The definitions of these
variables and a discussion of their properties can be found in Refs.\ \cite{ALEPH-qcdpaper} and
\cite{as_theory-uncertainties,NNLAreview}. 

The measurements have been carried out by the \ALEPH\ collaboration \cite{ALEPH-qcdpaper}
\footnote{The tables with numbers and uncertainties for all variables can be found at
{\tt http://aleph.web.cern.ch/aleph/QCD/alephqcd.html}.},
%and
%{\tt http://durpdg.dur.ac.uk/cgi-hepdata/hepreac/5765862} .} 
at centre-of-mass energies of 91.2, 133, 161, 172, 183, 189, 200 and 206 \GeV.
Earlier measurements and complementary data sets from the LEP experiments and 
from SLD can be found in Refs.~\cite{aleph,opal,l3,delphi,sld}.
The event-shape distri\-butions were computed using the reconstructed momenta and energies of charged and
neutral particles. 
The measurements have been corrected for detector effects, 
ie., the final distributions
correspond to the so-called particle (or hadron level). The particle
level is defined by stable hadrons with a lifetime longer than
$10^{-9}$ s after hadronisation and leptons according to the
definition given in \cite{aleph_mega}.
In addition, at LEP2 energies above the Z peak they were corrected for initial-state radiation effects.
 At energies above 133 GeV  backgrounds from
WW, ZZ and Z$\gamma^*$ were subtracted following the procedure
given in~\cite{ALEPH-qcdpaper}.

Backgrounds, mainly from W-pair production, were subtracted. The experimental uncertainties were estimated
by varying event and particle selection cuts. They are below 1\% at LEP1 and slightly larger at LEP2.
For further details we refer to Ref.\ \cite{ALEPH-qcdpaper}.

%%%%%%%%%%%%%%%%%%%%%%%%%%%%%%%%%%%%%%%%%%%%%%%%%
%%%%% Fits 
%%%%%%%%%%%%%%%%%%%%%%%%%%%%%%%%%%%%%%%%%%%%%%%%%%

\section{Determination of the strong coupling constant}
\label{sec:fits}

The coupling constant \as{}\ is
determined from a fit of the perturbative QCD predictions to measured
event-shape distributions. The procedure adopted here follows closely the one described in 
Ref.~\cite{ALEPH-qcdpaper}. The six event-shape variables used are $T$, $M_H$, $\bt$, $\bw$, $C$ and
$-\ln y_3$. 
The perturbative predictions for the distributions, as described in
section \ref{sec:theory}, are calculated to the same order of perturbation
theory for all of these variables. 
The data set collected by \ALEPH\ at the Z peak with high statistics allows for 
quantitative comparisons of different pQCD predictions and resulting fits 
for \as{}. The size of missing higher orders, which are inherently
difficult to assess, can be different for different variables. Therefore, a combination of 
measurements using several variables yields a better estimator of \as{}\ than using a single variable.
Furthermore, the spread of values of \as{}\ is an independent estimation of the
theoretical uncertainty. At centre-of-mass energies above the Z peak the statistical uncertainties are larger and background conditions are more
difficult compared to the peak data. Therefore a combination of measurements is particularly important for
those energies. We adopt the same combination procedure as described
in \cite{ALEPH-qcdpaper}, 
which is based on weighted averages and takes into account correlations between the event-shape variables. 
At energies above \Mz\  the measured statistical uncertainties are replaced 
by expected statistical errors, obtained from a large number of
simulated experiments. This avoids biases towards lower values of
\as{}\ in the case of small event statistics. The bias originates from larger weights of
downward fluctuating bins in the distributions compared to upward
fluctuations, as outlined in Ref.~\cite{ALEPH-qcdpaper}.
%% HS: following section from ALEPH replaced by ashorter version 
%A study using simulated distributions reveals that the fit
%procedure is systematically biased towards lower values of
%\as{}\ in the case of small event statistics, as encountered
%at 161 and 172 GeV. This bias originates from larger weights of
%downward fluctuating bins in the distributions compared to upward
%fluctuations. It is overcome by replacing the measured statistical
%uncertainties of a distribution by the expected statistical
%errors. The expected uncertainties are obtained from a large
%number of simulated experiments, each of the same sample size as
%the real data. The root mean square in each bin of the Monte Carlo
%distributions is used in the fit procedure as statistical error.
%This is done for all variables at all energy points above \Mz.

%% HS again a bit shortened and diluted ALEPH-style
Event-shape distributions are fit in a central region of
three-jet production, where a good perturbative description is
available. The fit range is
placed inside the region where hadronisation and detector
corrections are below 25$\%$ and the signal-to-background ratio at LEP2 is
above one.  At the higher LEP2 energies the good perturbative description extends 
further into the two-jet region, while in the four-jet region 
the background becomes large. Thus the fit range is selected as a result 
of an iterative procedure balancing theoretical, experimental and 
statistical uncertainties.  
The data are corrected for detector
effects, for background from four-fermion processes and for the
residual ISR contribution.
The background from WW events increases with energy, and after
subtraction the content of some bins of the distribution becomes negative. For
this reason the fit range is restricted to a region with good
signal-to-background ratio.

Since very recently distributions of infrared- and collinear-safe observables at the
parton level can be computed in perturbative QCD to third order
in \as{},  as described in section \ref{sec:theory} and in more detail 
in Ref.~\cite{eerad3}. Here we concentrate on fits of NNLO 
predictions and compare them to pure NLO and matched NLO+NLLA predictions 
as used in the analysis of Ref.\ \cite{ALEPH-qcdpaper}. 
The nominal value for the renormalisation scale $x_\mu = \mu/Q$ is
unity. All these calculations neglect quark masses. Quark mass
effects~\cite{sherpa} are relevant for the b quark at $\sqrt{s}=\Mz$, where the
fraction of ${\rm b\overline{b}}$ events is large, while
$Q$ is still moderate. NLO calculations including a quark
mass indicate that the expected change in \as{}\ is of the
order of $1\%$ at \Mz\ \cite{aleph_mb}. The effect is scaling
with $M_{\rm b}^2/Q^2$ and decreases to $0.2$-$0.3\%$ at 200 GeV.
Mass corrections were computed to second order using the matrix
elements of \cite{quarkmass}. A pole b-quark mass $M_{\rm b}$ = 5\,
\GeVcc\ was used and Standard Model values were taken for the
fraction of ${\rm b\overline{b}}$ events. 
It is worth noting that, since these mass effects are known at NLO only, 
the NNLO predictions can be corrected only in a partial and not 
fully consistent manner. 
\begin{figure}[h]
\vspace{-4mm}
\begin{tabular}{cc}
\hspace*{-0.5cm}\includegraphics[width=8cm]{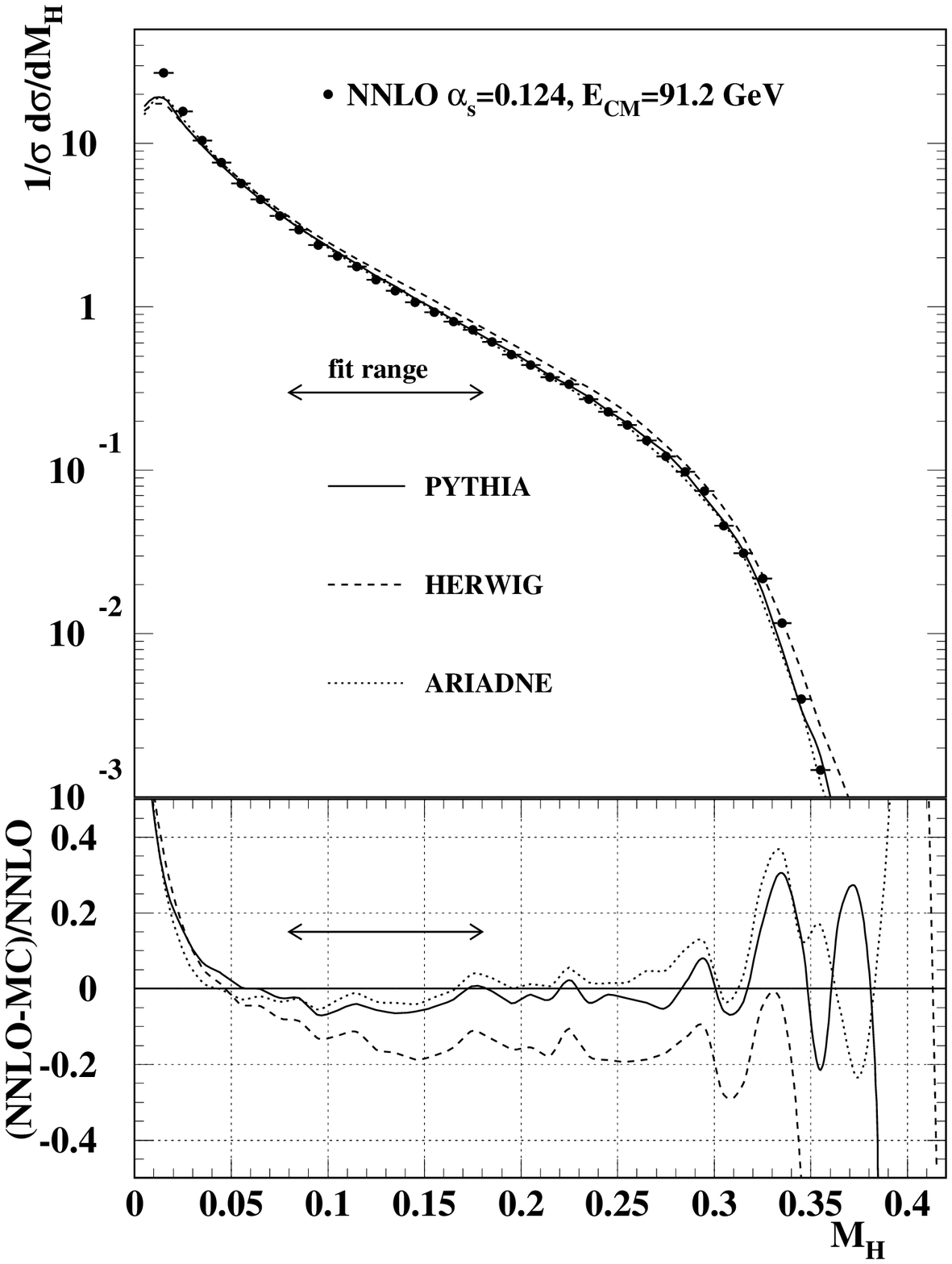} &
\hspace*{-0.8cm}\includegraphics[width=8cm]{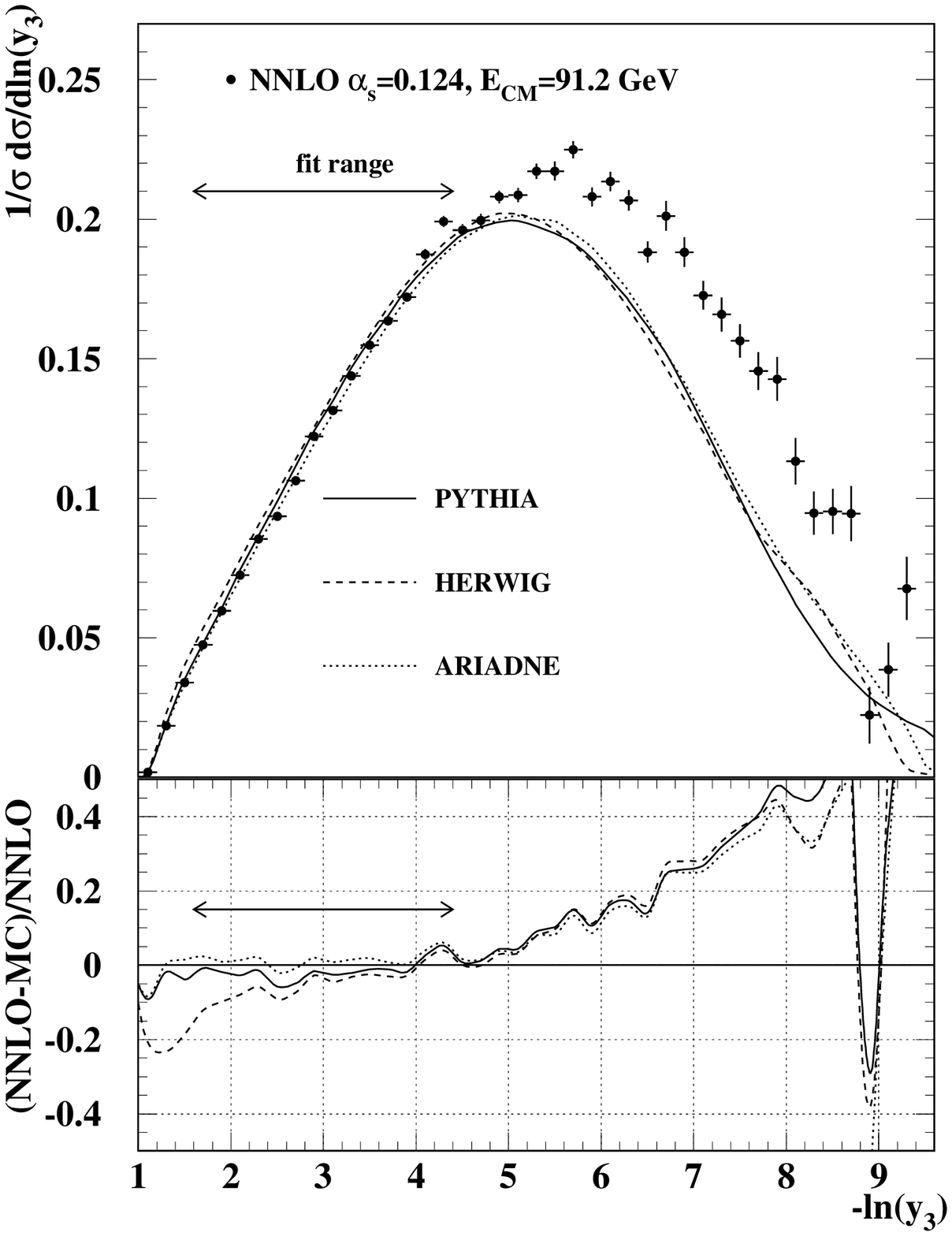}
\end{tabular}
\caption{\small Comparison of the parton-level distributions predicted by the 
generators 
and the NNLO calculation for the heavy jet mass (left) and -ln(y3) (right). }
\protect\label{fig:hp}
\end{figure}

The perturbative QCD prediction is corrected for hadronisation and resonance
decays  by means of a transition matrix,
which is computed with the Monte Carlo
generators
PYTHIA, HERWIG and ARIADNE, all tuned to global hadronic observables at
$\Mz$ \cite{aleph_mega}. The parton level is defined by the quarks and gluons
present at
the end of the parton shower in PYTHIA and HERWIG and the partons resulting
from the
color dipole radiation in ARIADNE. Albeit the generators' limitation to
leading-order QCD,
a reasonable agreement is observed between the generator parton level and
the NNLO prediction inside the fit range within a few percent,
 as shown in Fig.~\ref{fig:hp}. 
Corrected measurements of event-shape distributions are compared 
to the theoretical calculation at particle level.

\begin{figure}[h]
\begin{tabular}{cc}
\hspace*{-0.5cm}\includegraphics[width=8cm]{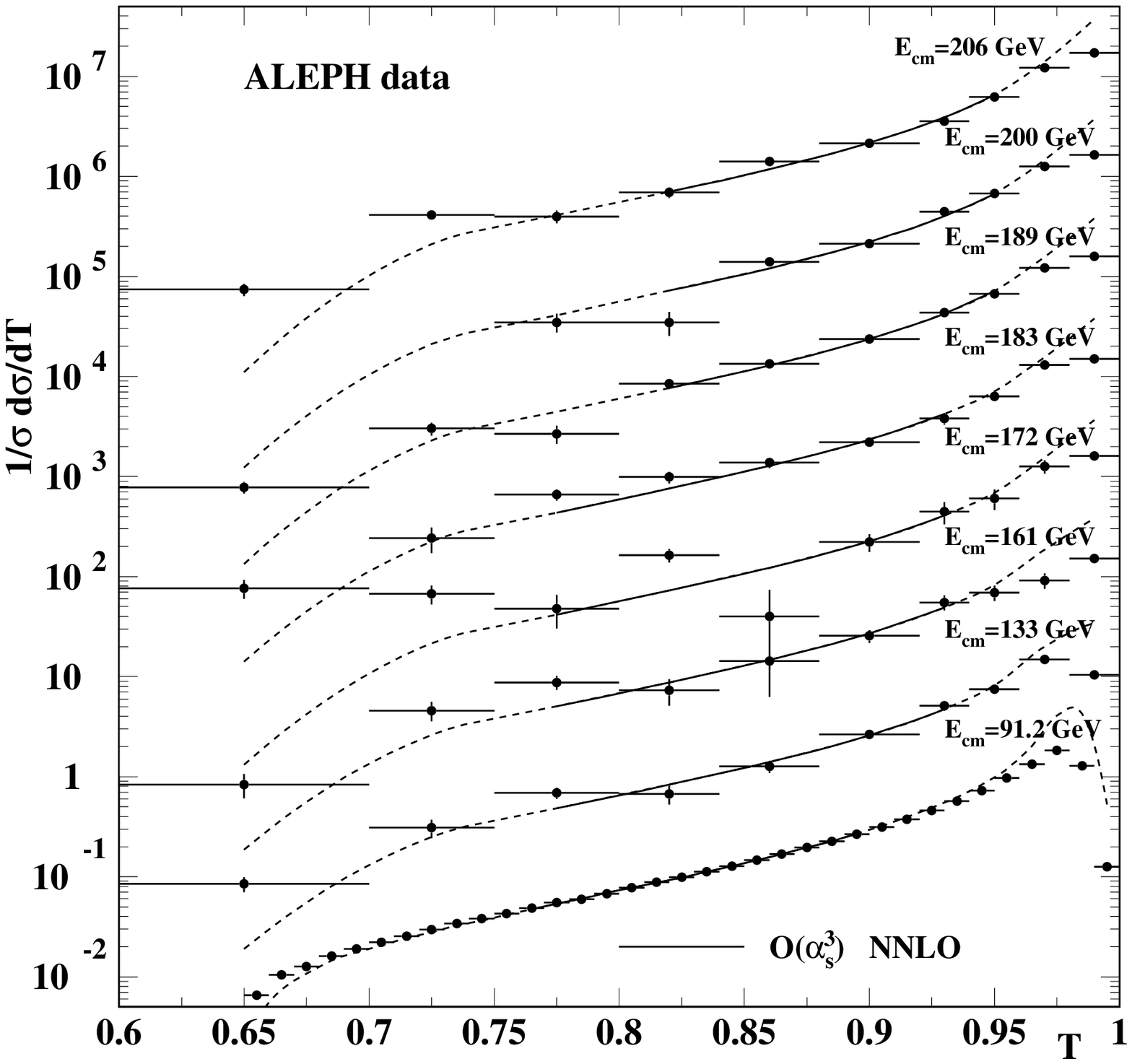} &
\hspace*{-0.8cm}\includegraphics[width=8cm]{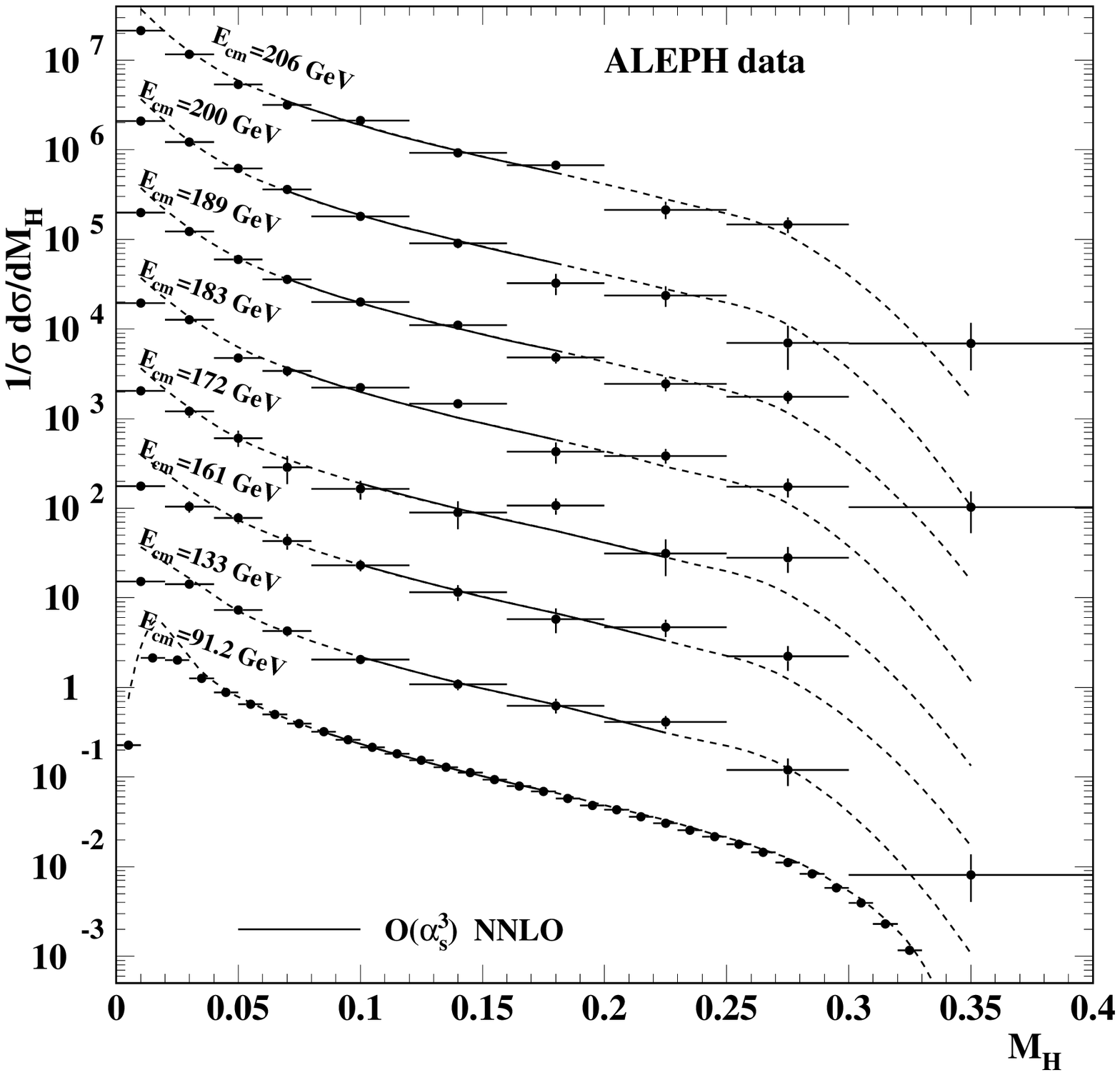} \\
\hspace*{-0.5cm}\includegraphics[width=8cm]{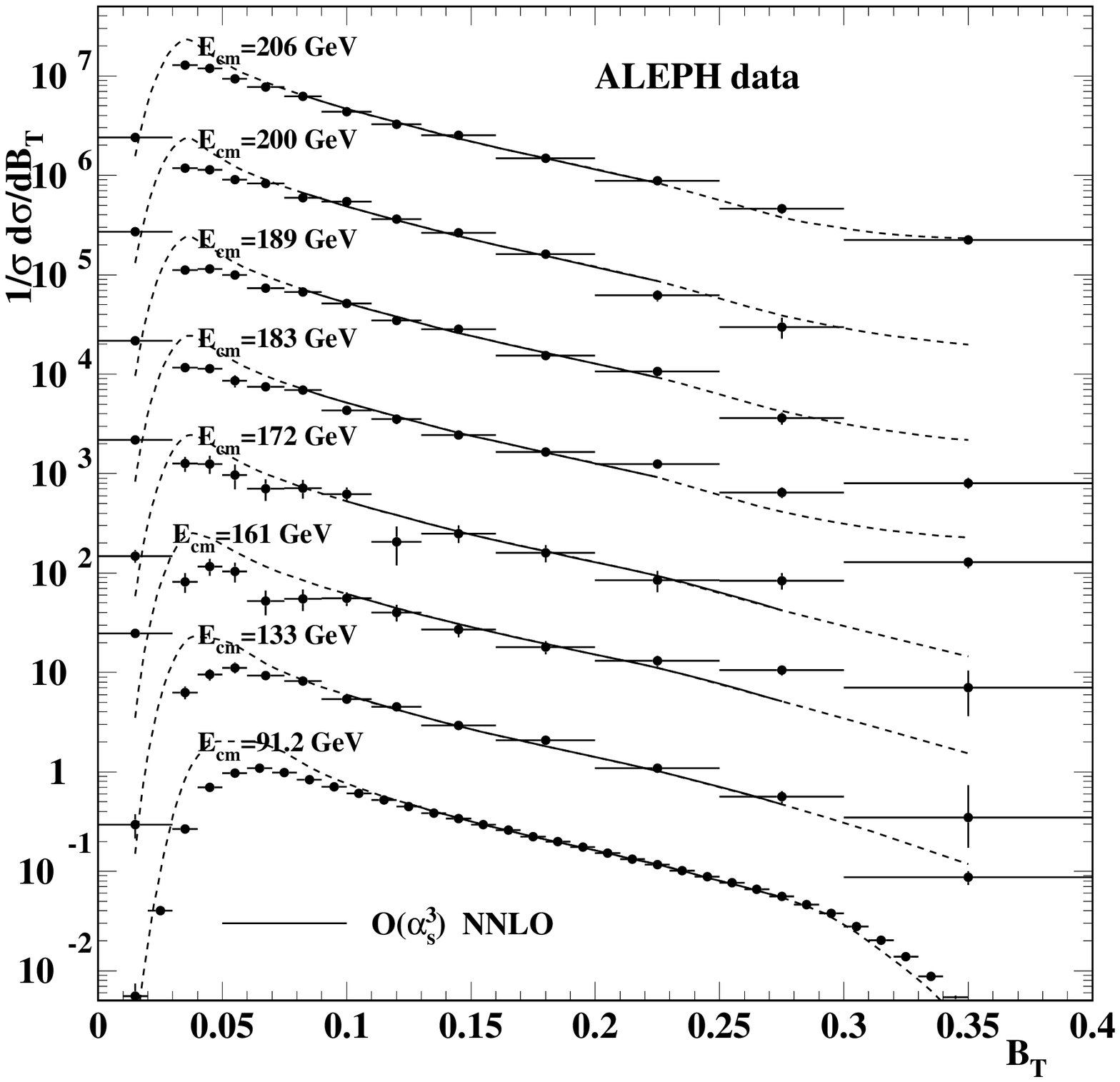} &
\hspace*{-0.8cm}\includegraphics[width=8cm]{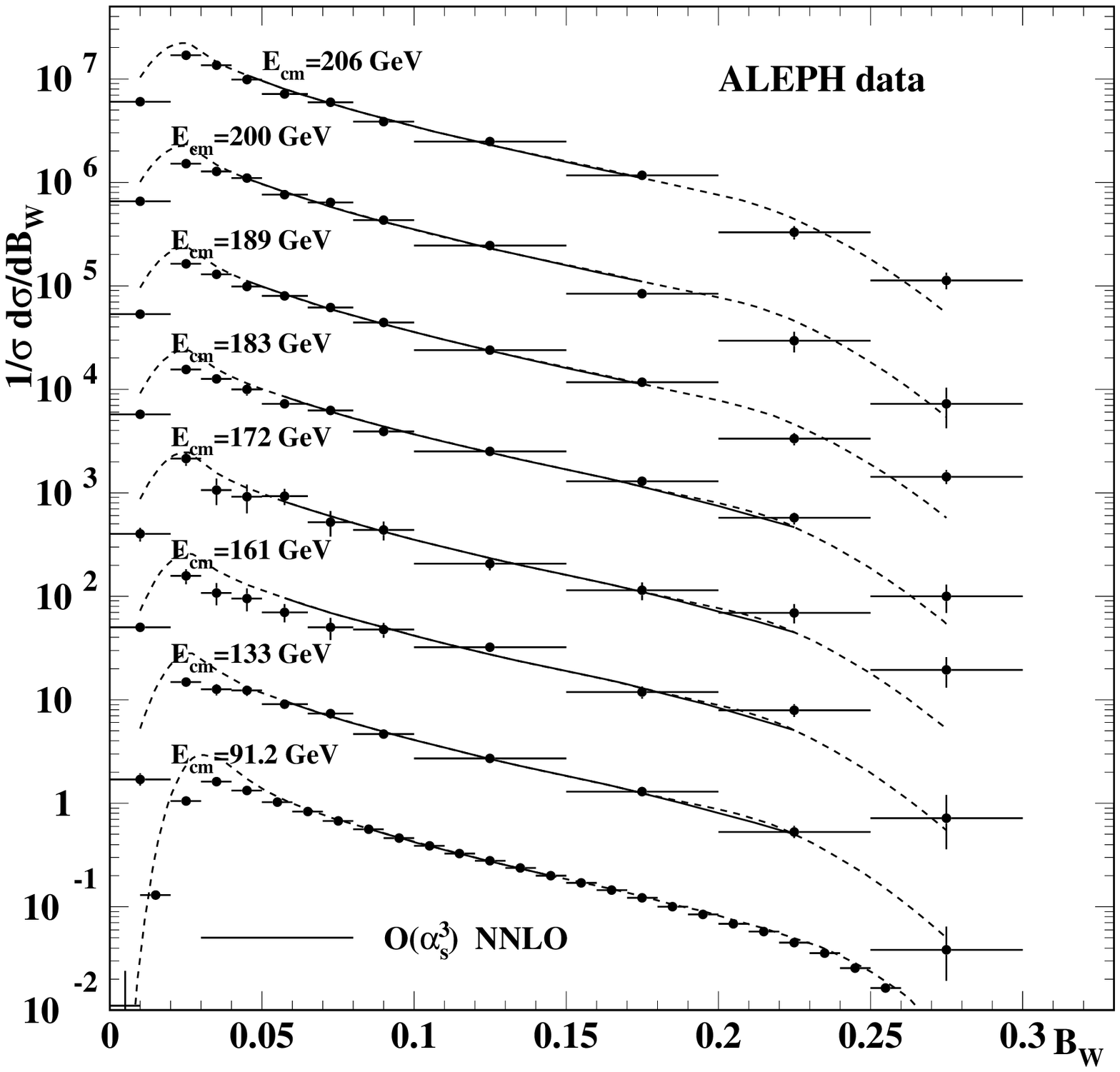} \\
\end{tabular}
\caption{\small Distributions measured by \ALEPH, after correction
for backgrounds and detector effects, of thrust,  heavy jet mass, total and wide jet broadening at energies between
91.2 and 206 GeV together with the fitted NNLO QCD predictions.
The error bars correspond to statistical uncertainties.
The fit ranges
cover the central regions indicated by the solid curves,
the theoretical predictions extrapolate well outside the fit ranges, as shown by the
dotted curves. The plotted
distributions are scaled by arbitrary factors for presentation.}
\protect\label{fig:alphas_fits1}
\end{figure}

\begin{figure}[h!]
\begin{tabular}{cc}
\hspace*{-0.5cm}\includegraphics[width=8.0cm]{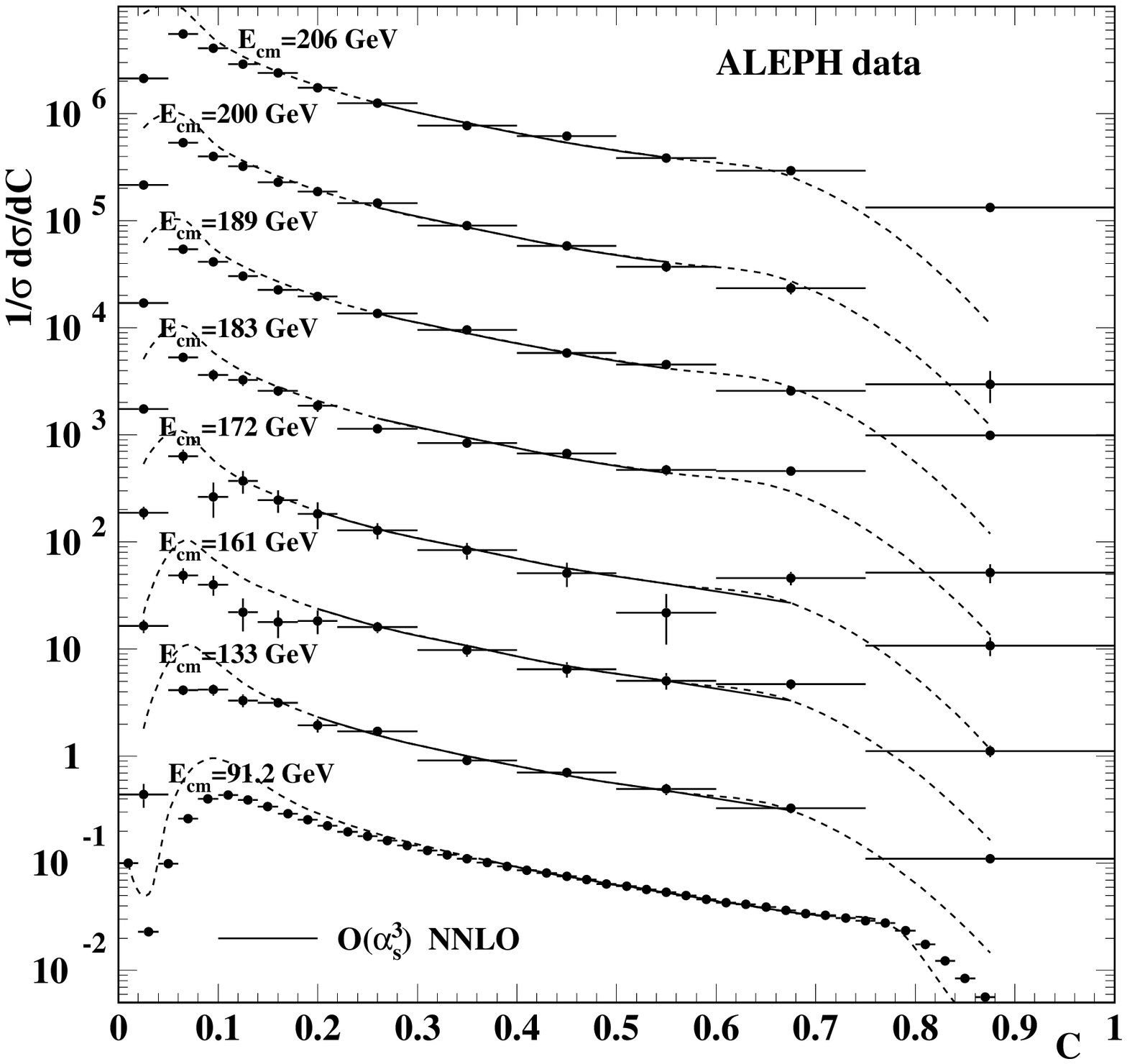} &
\hspace*{-0.8cm}\includegraphics[width=8.0cm]{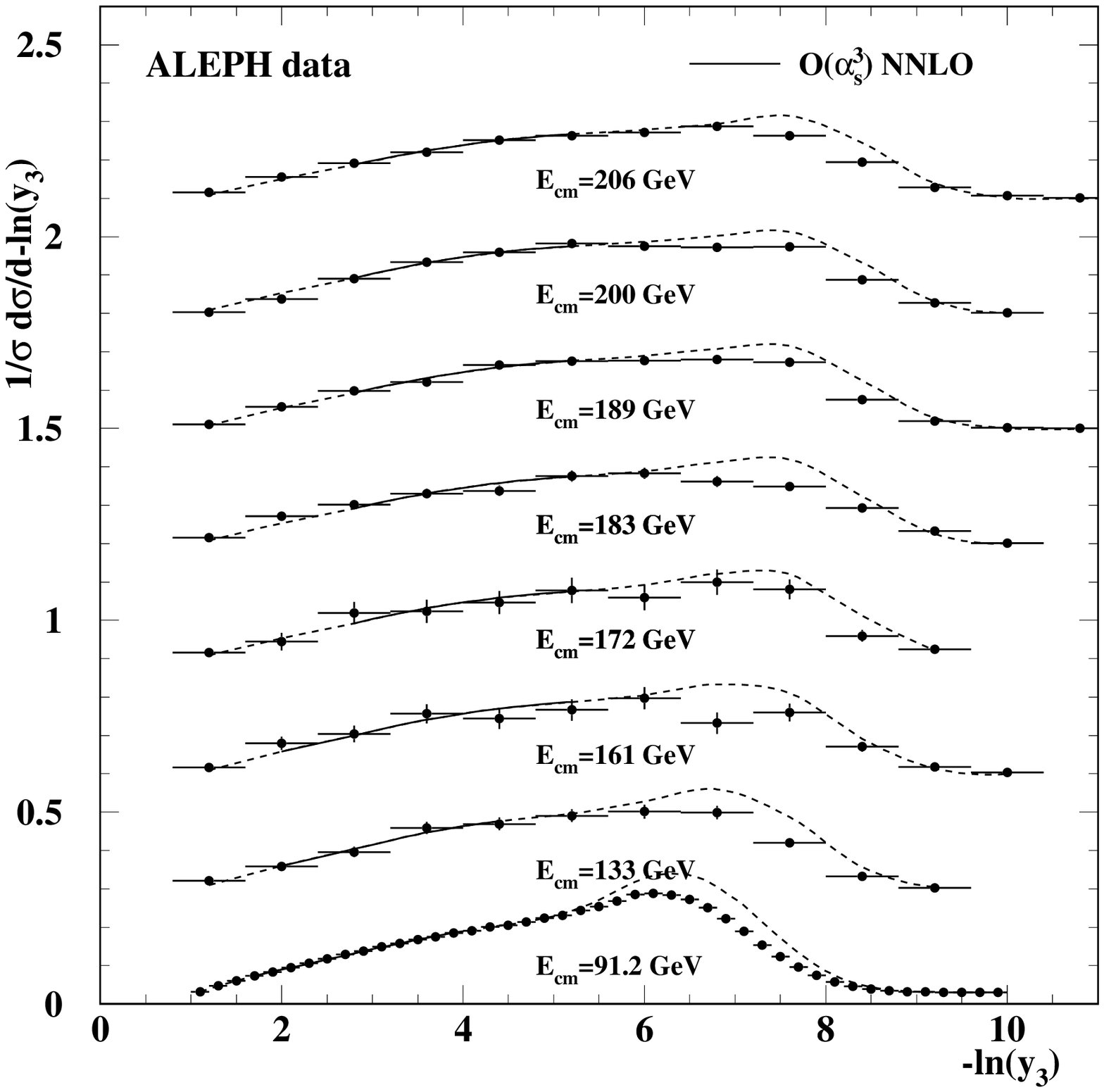} 
\end{tabular}
\caption{\small Distributions measured by \ALEPH, after correction
for backgrounds and detector effects, of C-parameter
and  the three-jet transition variable at energies between 91.2
and 206 GeV together with the fitted NNLO QCD predictions.
The error bars correspond to statistical uncertainties.
The fit ranges
cover the central regions indicated by the solid curves,
the theoretical predictions extrapolate well outside the fit ranges, as
shown by the dotted curves. The plotted
distributions are scaled by arbitrary factors for presentation.}
\protect\label{fig:alphas_fits2}
\end{figure}

The value of \as{}\ is determined at each energy using a binned
least-squares fit. The fit programs of  Ref.\ \cite{ALEPH-qcdpaper} 
have 
 been extended to incorporate the NNLO calculations. It has been verified 
that at NLO+NLLA and using the fit range 
advocated by ALEPH the results on $\alpha_s$ 
reported in   Ref.\ \cite{ALEPH-qcdpaper} are reproduced. 
Only statistical uncertainties arising from 
the limited number of observed events, from the number of simulated  
events used to calculate hadronisation and detector corrections and 
from the integration procedure of the NNLO coefficient functions  
are included in
the $\chi^2$ of the fit. Its quality is good for all variables
at all energies. 
Nominal results, based on (\ref{eq:NNLO}) and using the fitted values of 
\as{}\, are shown in
Figs.~\ref{fig:alphas_fits1} and \ref{fig:alphas_fits2} together
with the measured distributions. The resulting measurements of
$\as{}(Q)$ are given in Table~\ref{tab:indiv1} for 91.2 to 172
GeV and in Table~\ref{tab:indiv2} for 183 to 206 GeV.

\begin{figure}[th]
\begin{tabular}{lr}
\hspace*{-0.5cm}\includegraphics[width=7.7cm]{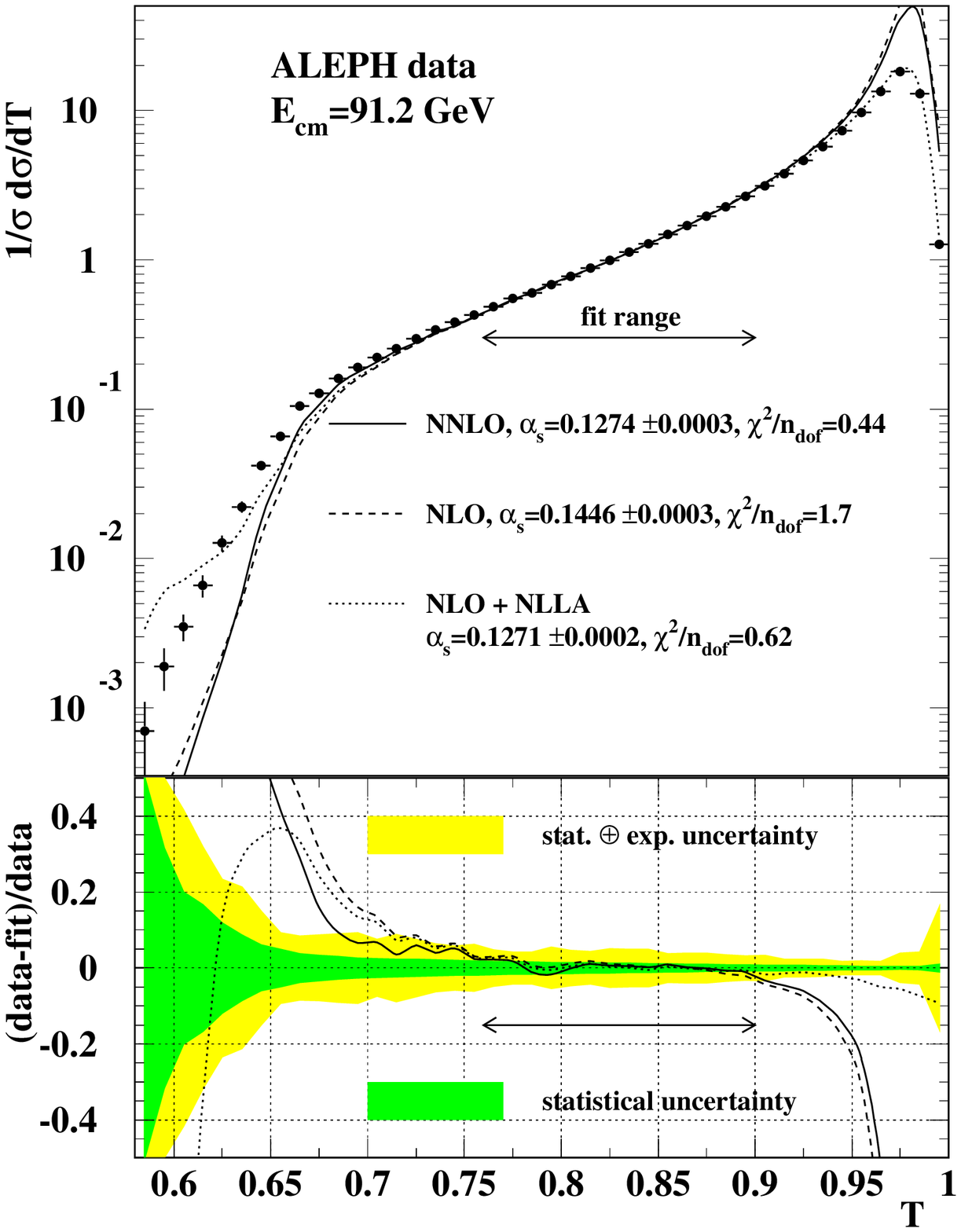} &
\hspace*{-0.5cm}\includegraphics[width=7.7cm]{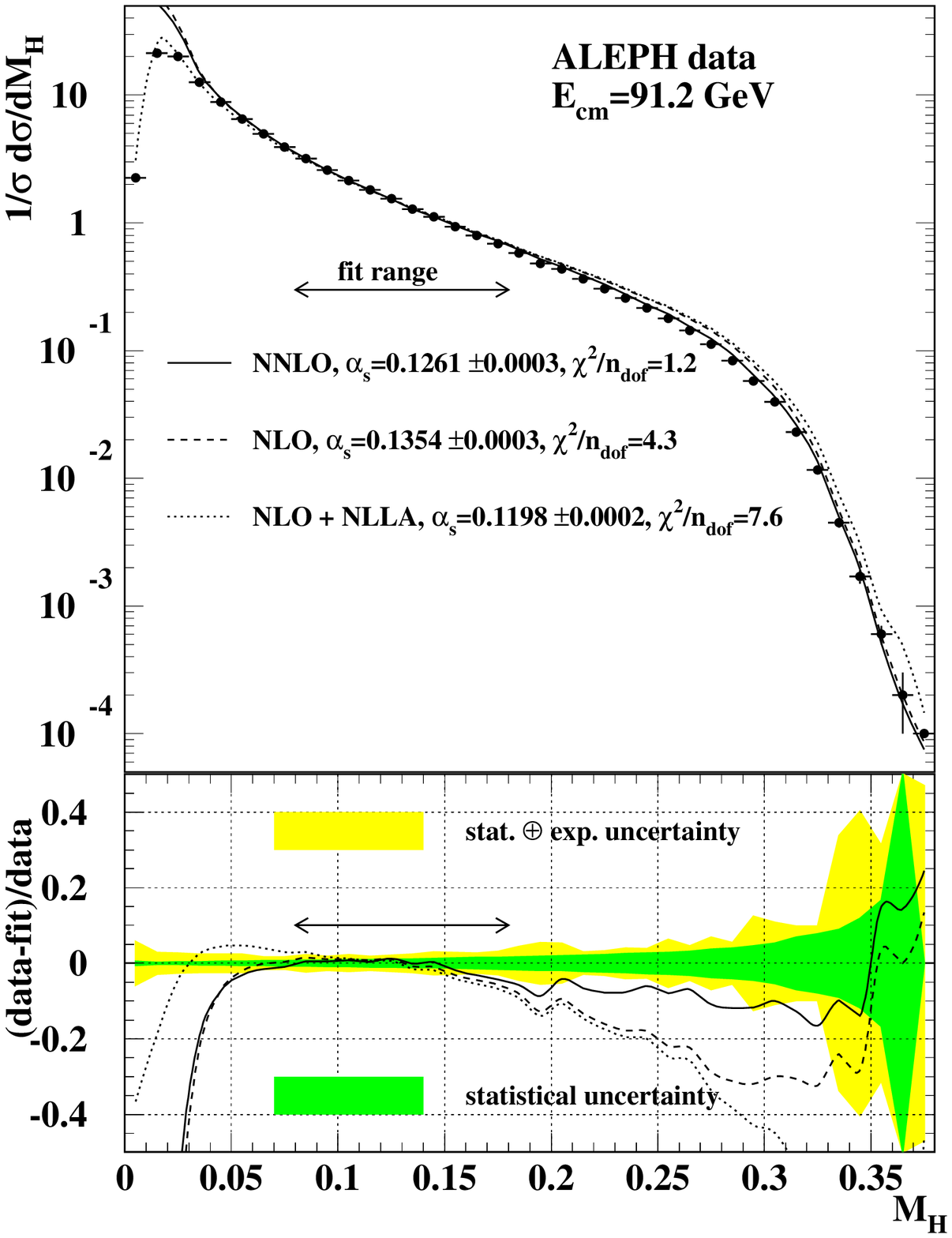} \\[-6mm]
\hspace*{-0.5cm}\includegraphics[width=7.7cm]{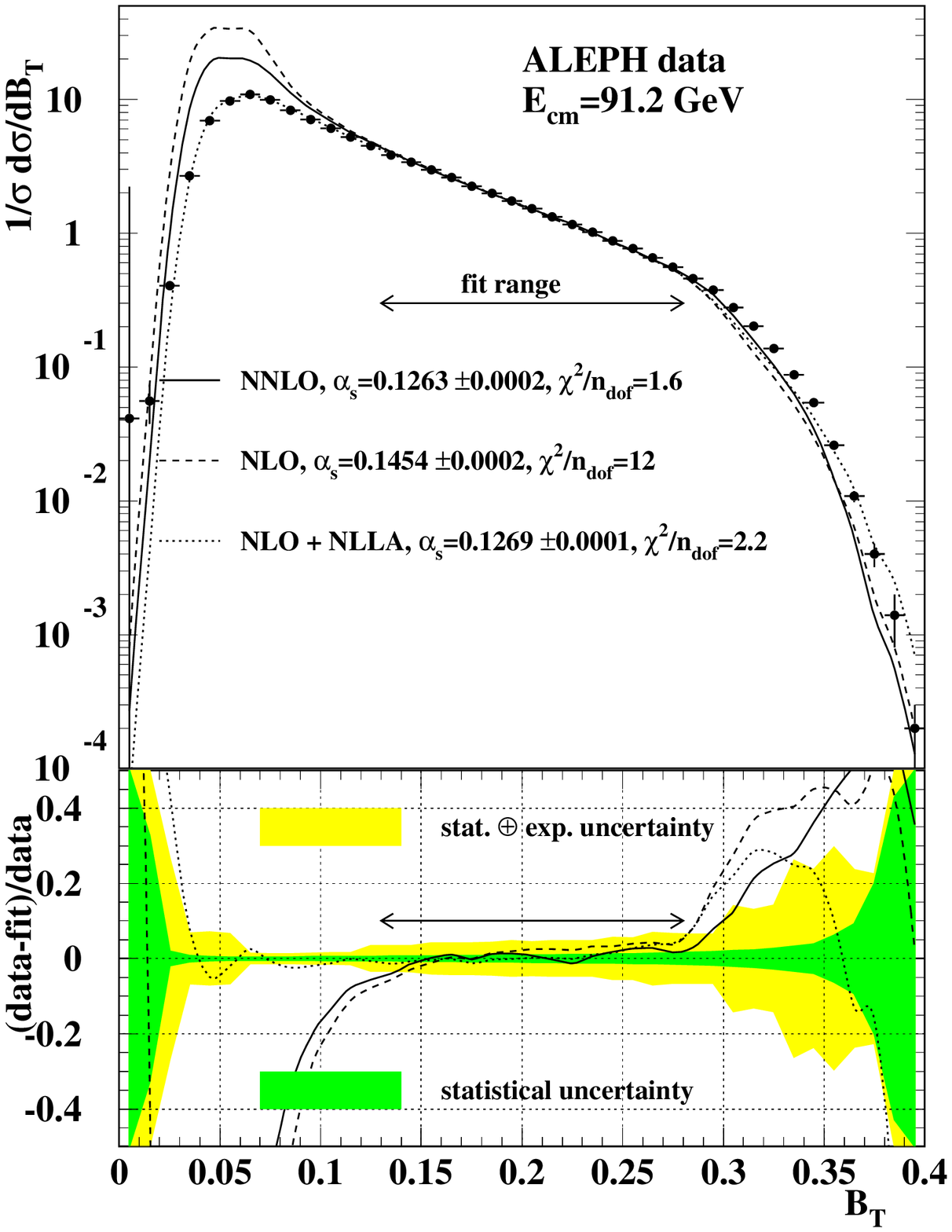} &
\hspace*{-0.5cm}\includegraphics[width=7.7cm]{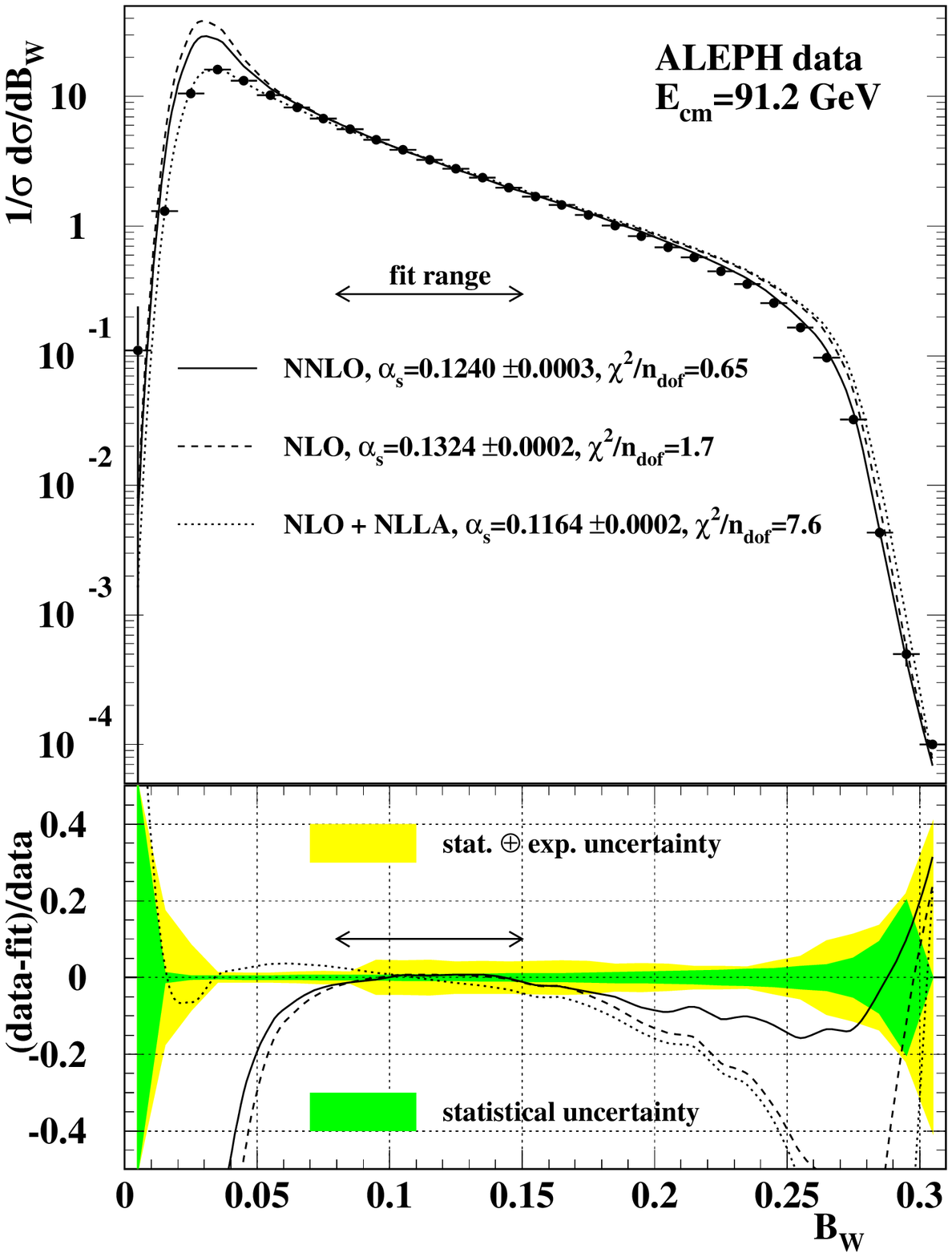} 
\end{tabular}
\caption{\small Distributions measured by \ALEPH\ at LEP1, after correction
for backgrounds and detector effects, of thrust,
heavy jet mass, total and wide jet broadening. Fitted QCD predictions at different orders of perturbation theory
are overlaid. The lower insets show a relative comparison of data and QCD fits.
}
\protect\label{fig:comp_nlo_nnlo1}
\end{figure}

\begin{figure}[th]
\begin{tabular}{lr}
\hspace*{-0.5cm}\includegraphics[width=7.7cm]{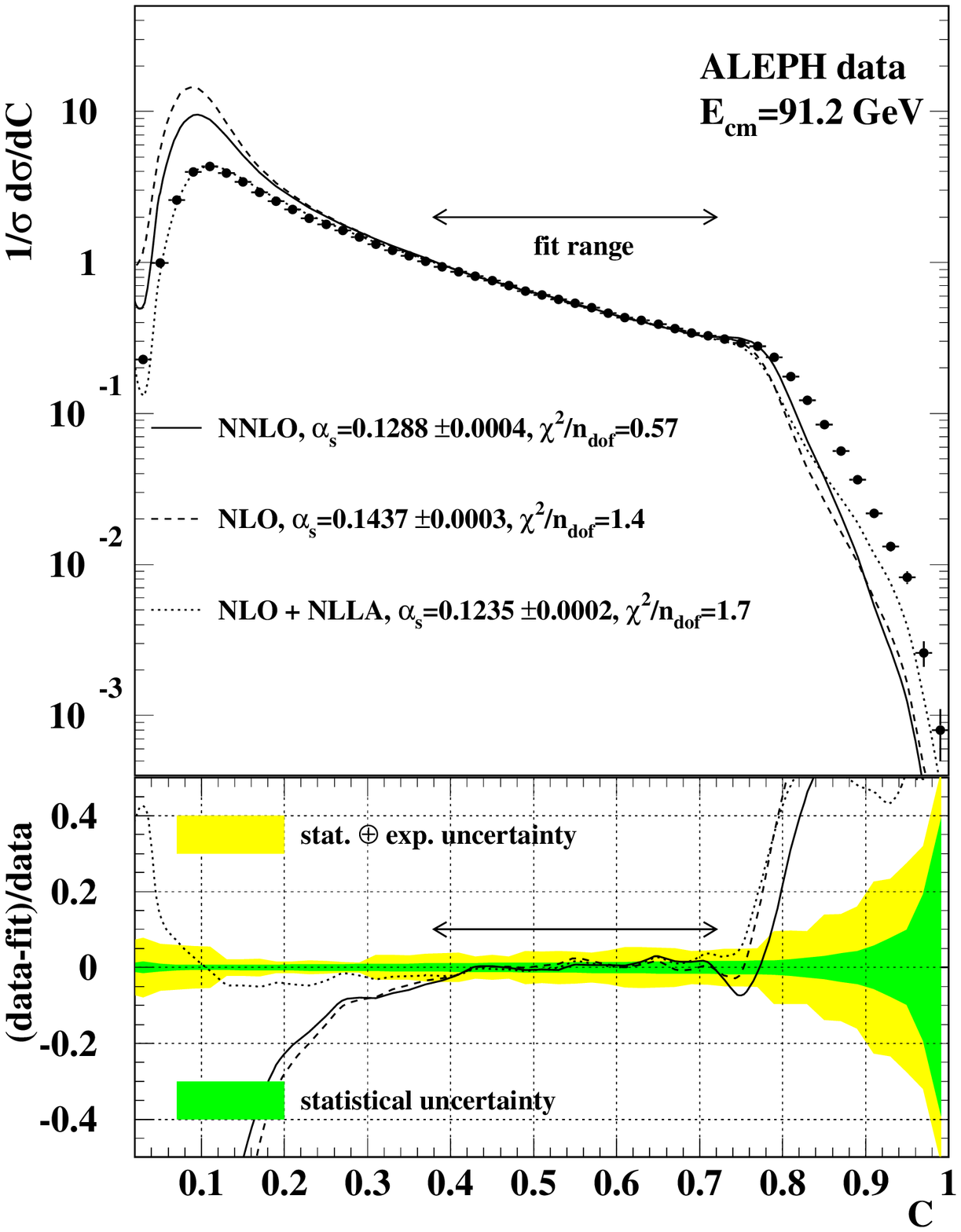} &
\hspace*{-0.5cm}\includegraphics[width=7.7cm]{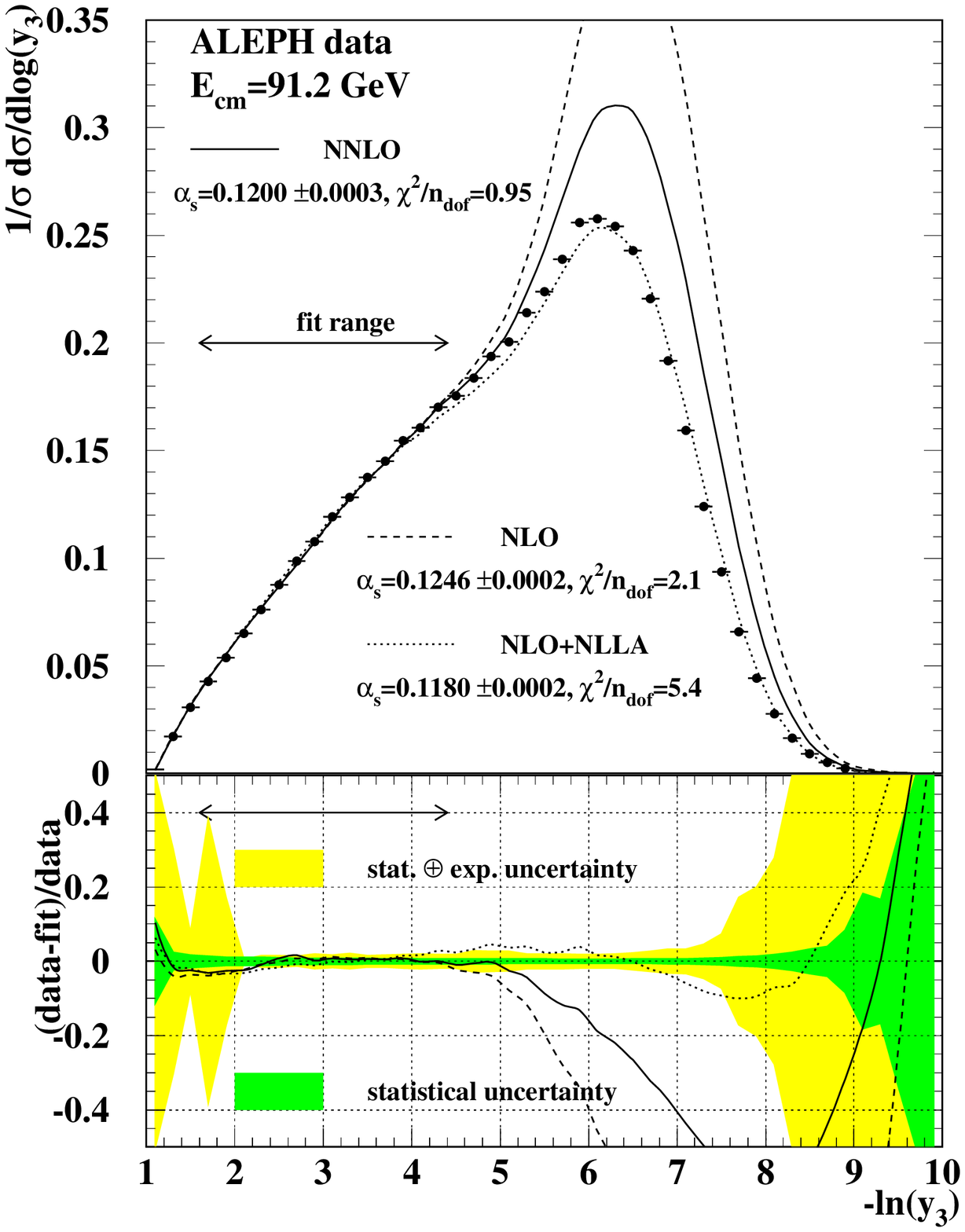} 
\end{tabular}
\caption{Distributions measured by \ALEPH\ at LEP1, after correction
for backgrounds and detector effects, of C-parameter
and the three-jet transition variable. Fitted QCD predictions at different orders of perturbation theory
are overlaid. The lower insets show a relative comparison of data and QCD fits.
}
\protect\label{fig:comp_nlo_nnlo2}
\end{figure}

Furthermore, in Figs.~\ref{fig:comp_nlo_nnlo1} and \ref{fig:comp_nlo_nnlo2}  
and in 
Table~\ref{tab:comp_nlo_nnlo} we compare the results found when fitting
different perturbative predictions to the high-precision 
data at \Mz. In particular we compare the NNLO fits to those obtained with NLO and
NLO+NLLA predictions. A detailed
description of the latter is given in Ref.~\cite{ALEPH-qcdpaper}.
As shown in Figs.~\ref{fig:comp_nlo_nnlo1} and
\ref{fig:comp_nlo_nnlo2} a good description 
by the NNLO predictions is found for an extended range over the perturbative region compared 
to NLO. However, in the two-jet region the resummed prediction 
still yields a better result. Therefore, compared to Ref.~\cite{ALEPH-qcdpaper}, 
the fit range for NNLO and NLO is slightly shifted into the perturbative region.   
We observe a clear improvement in the fit quality when going to
NNLO accuracy. Compared to NLO the value of \as{} is lowered 
by about 10\%, but still higher than for NLO+NLLA. 
Systematic theoretical and experimental uncertainties are
discussed in the following section.

%% Results, 13.11.2007, central results, stat, th and exp errors.
\renewcommand{\arraystretch}{1.1}
\renewcommand{\textfraction}{0.0001}
\begin{table}[h]
\caption[Individual results] {\label{tab:indiv1}{\small Results for
$\alpha_s(Q)$ as obtained from NNLO fits to distributions of
event-shape variables at $Q = \sqrt{s} =$ 91.2, 133, 161 and 172 GeV.}}
\begin{center}
\begin{tabular}{|l|c|c|c|c|c|c|}\hline
\multicolumn{7}{|c|}{ $Q$ = 91.2 GeV} \\ \hline 
variable        & $T$   & $-\ln y_3$ & $M_H$ & $C$ & $B_W$ & $B_T$  \\ \hline
$\alpha_s$      & 0.1274 & 0.1200 & 0.1261 & 0.1288 & 0.1240 & 0.1263 \\
stat. error     & 0.0003 & 0.0003 & 0.0003 & 0.0004 & 0.0003 & 0.0002 \\
exp. error      & 0.0008 & 0.0010 & 0.0011 & 0.0008 & 0.0007 & 0.0007 \\
pert. error     & 0.0042 & 0.0039 & 0.0029 & 0.0039 & 0.0029 & 0.0047 \\
hadr. error     & 0.0019 & 0.0016 & 0.0044 & 0.0017 & 0.0018 & 0.0018 \\
total error     & 0.0047 & 0.0043 & 0.0054 & 0.0043 & 0.0035 & 0.0051 \\ \hline
fit range       & 0.76-0.90 & 1.6-4.4 & 0.08-0.18 & 0.38-0.72 & 0.08-0.15 & 0.13-0.28 \\ \hline
\multicolumn{7}{|c|}{$Q$ = 133 GeV} \\ \hline 
variable        &  $T$   & $-\ln y_3$ & $M_H$ & $C$ & $B_W$ & $B_T$  \\ \hline
$\alpha_s$      & 0.1197 & 0.1198 & 0.1216 & 0.1208 & 0.1211 & 0.1177 \\
stat. error     & 0.0035 & 0.0046 & 0.0042 & 0.0030 & 0.0030 & 0.0026 \\
exp. error      & 0.0010 & 0.0007 & 0.0011 & 0.0011 & 0.0005 & 0.0012 \\
pert. error     & 0.0034 & 0.0034 & 0.0025 & 0.0032 & 0.0022 & 0.0038 \\
hadr. error     & 0.0014 & 0.0009 & 0.0028 & 0.0012 & 0.0011 & 0.0014 \\
total error     & 0.0052 & 0.0058 & 0.0058 & 0.0047 & 0.0039 & 0.0049 \\
fit range       & 0.75-0.94 & 1.6-4.8 & 0.06-0.25 & 0.22-0.75 & 0.05-0.25 & 0.09-0.30 \\ \hline 
\multicolumn{7}{|c|}{$Q$ = 161 GeV} \\ \hline 
variable        &  $T$   & $-\ln y_3$ & $M_H$ & $C$ & $B_W$ & $B_T$  \\ \hline
$\alpha_s$      & 0.1239 & 0.1146 & 0.1267 & 0.1259 & 0.1223 & 0.1231 \\
stat. error     & 0.0054 & 0.0074 & 0.0078 & 0.0046 & 0.0046 & 0.0041 \\
exp. error      & 0.0010 & 0.0007 & 0.0011 & 0.0011 & 0.0005 & 0.0012 \\
pert. error     & 0.0031 & 0.0032 & 0.0023 & 0.0029 & 0.0020 & 0.0035 \\
hadr. error     & 0.0012 & 0.0006 & 0.0022 & 0.0011 & 0.0008 & 0.0013 \\
total error     & 0.0064 & 0.0081 & 0.0085 & 0.0056 & 0.0051 & 0.0056 \\\hline
fit range       & 0.75-0.94 & 1.2-5.6 & 0.06-0.25 & 0.22-0.75 & 0.05-0.25 & 0.03-0.30 \\ \hline
\multicolumn{7}{|c|}{ $Q$ = 172 GeV} \\ \hline 
variable        & $T$   & $-\ln y_3$ & $M_H$ & $C$ & $B_W$ & $B_T$  \\ \hline
$\alpha_s$      & 0.1101 & 0.1079 & 0.1095 & 0.1104 & 0.1083 & 0.1125 \\
stat. error     & 0.0072 & 0.0089 & 0.0087 & 0.0064 & 0.0059 & 0.0056 \\
exp. error      & 0.0010 & 0.0009 & 0.0011 & 0.0011 & 0.0005 & 0.0012 \\
pert. error     & 0.0030 & 0.0032 & 0.0022 & 0.0028 & 0.0019 & 0.0033 \\
hadr. error     & 0.0012 & 0.0005 & 0.0021 & 0.0010 & 0.0007 & 0.0012 \\
total error     & 0.0079 & 0.0093 & 0.0092 & 0.0069 & 0.0063 & 0.0067 \\\hline
fit range     & 0.80-0.96 & 1.2-5.6 & 0.06-0.25 & 0.22-0.75 & 0.05-0.25 & 0.09-0.30 \\ \hline
\end{tabular}
\end{center}
\end{table}

 %% Results, 07.11.2007, central results, stat, th and exp errors.
\renewcommand{\arraystretch}{1.1}
\renewcommand{\textfraction}{0.0001}
\begin{table}[h]
\caption[Individual results] {\label{tab:indiv2}{\small Results for
$\alpha_s(Q)$ as obtained from NNLO fits to distributions of
event-shape variables at $Q = \sqrt{s} = $183, 189, 200 and 206 GeV.}}
\begin{center}
\begin{tabular}{|l|c|c|c|c|c|c|}\hline
\multicolumn{7}{|c|}{ $Q$ = 183 GeV} \\ \hline 
variable          & $T$   & $-\ln y_3$ & $M_H$ & $C$ & $B_W$ & $B_T$  \\ \hline
$\alpha_s$        & 0.1132 & 0.1070 & 0.1133 & 0.1160 & 0.1114 & 0.1117 \\
stat. error       & 0.0032 & 0.0041 & 0.0039 & 0.0029 & 0.0027 & 0.0027 \\
exp. error        & 0.0011 & 0.0009 & 0.0012 & 0.0014 & 0.0005 & 0.0012 \\
pert. error       & 0.0029 & 0.0031 & 0.0022 & 0.0027 & 0.0019 & 0.0033 \\
hadr. error       & 0.0011 & 0.0004 & 0.0019 & 0.0010 & 0.0007 & 0.0012 \\
total error       & 0.0046 & 0.0052 & 0.050 & 0.0043 & 0.0033 & 0.0046 \\ \hline
fit range     & 0.75-0.94 & 1.2-5.6 & 0.06-0.25 & 0.22-0.75 & 0.05-0.25 & 0.09-0.30 \\ \hline 
\multicolumn{7}{|c|}{$Q$ = 189 GeV} \\ \hline 
variable          &  $T$   & $-\ln y_3$ & $M_H$ & $C$ & $B_W$ & $B_T$  \\ \hline
$\alpha_s$        & 0.1140 & 0.1077 & 0.1118 & 0.1122 & 0.1093 & 0.1124 \\
stat. error       & 0.0020 & 0.0025 & 0.0025 & 0.0017 & 0.0016 & 0.0015 \\
exp. error        & 0.0010 & 0.0007 & 0.0013 & 0.0011 & 0.0005 & 0.0013 \\
pert. error       & 0.0028 & 0.0031 & 0.0021 & 0.0026 & 0.0018 & 0.0030 \\
hadr. error       & 0.0011 & 0.0004 & 0.0018 & 0.0009 & 0.0006 & 0.0012 \\
total error       & 0.0037 & 0.0041 & 0.0040 & 0.0034 & 0.0025 & 0.0038 \\ \hline
fit range     & 0.80-0.96 & 1.6-5.6 & 0.04-0.20 & 0.18-0.60 & 0.04-0.20 & 0.075-0.25 \\ \hline 
\multicolumn{7}{|c|}{$Q$ = 200 GeV} \\ \hline 
variable          &  $T$   & $-\ln y_3$ & $M_H$ & $C$ & $B_W$ & $B_T$  \\ \hline
$\alpha_s$        & 0.1094 & 0.1071 & 0.1079 & 0.1105 & 0.1071 & 0.1082 \\
stat. error       & 0.0022 & 0.0026 & 0.0027 & 0.0018 & 0.0017 & 0.0017 \\
exp. error        & 0.0010 & 0.0007 & 0.0014 & 0.0011 & 0.0005 & 0.0013 \\
pert. error       & 0.0027 & 0.0031 & 0.0021 & 0.0026 & 0.0018 & 0.0029 \\
hadr. error       & 0.0010 & 0.0003 & 0.0017 & 0.0009 & 0.0006 & 0.0011 \\
total error       & 0.0037 & 0.0041 & 0.0041 & 0.0035 & 0.0026 & 0.0038 \\ \hline
fit range     & 0.80-0.96 & 1.6-5.6 & 0.04-0.20 & 0.18-0.60 & 0.04-0.20 & 0.075-0.25 \\ \hline 
\multicolumn{7}{|c|}{$Q$ = 206 GeV} \\ \hline 
variable          &  $T$   & $-\ln y_3$ & $M_H$ & $C$ & $B_W$ & $B_T$  \\ \hline
$\alpha_s$        & 0.1075 & 0.1023 & 0.1093 & 0.1064 & 0.1066 & 0.1056 \\
stat. error       & 0.0021 & 0.0026 & 0.0025 & 0.0017 & 0.0016 & 0.0016 \\
exp. error        & 0.0010 & 0.0007 & 0.0011 & 0.0011 & 0.0005 & 0.0012 \\
pert. error       & 0.0026 & 0.0031 & 0.0021 & 0.0025 & 0.0017 & 0.0029 \\
hadr. error       & 0.0010 & 0.0003 & 0.0016 & 0.0009 & 0.0005 & 0.0011 \\
total error       & 0.0037 & 0.0041 & 0.0038 & 0.0034 & 0.0025 & 0.0037 \\ \hline
fit range     & 0.80-0.96 & 1.6-5.6 & 0.04-0.20 & 0.18-0.60 & 0.04-0.20 & 0.075-0.25 \\ \hline
\end{tabular}
\end{center}
\end{table}


 %% Results, 07.11.2007, central results, stat, th and exp errors.
 \begin{table}[h]
\caption[Some fits] {\label{tab:comp_nlo_nnlo}{\small Various fit results at a centre-of-mass energy of 91.2 GeV, using 
different predictions of perturbative QCD, with the renormalisation
scale fixed to $\mu = \Mz$.}}
\begin{center}
\begin{tabular}{|l|c|c|c|c|c|c|}\hline
          & $T$         & $C$         & $M_H$       & $B_W$       & $B_T$ & $-\ln y_3$\\ \hline
fit range & 0.76 - 0.90 & 0.38 - 0.72 & 0.08 - 0.18 & 0.08 - 0.15 & 0.13 - 0.28 & 1.6-4.4 \\ \hline
NNLO      & 0.1274      & 0.1288      & 0.1261      & 0.1240      & 0.1263 & 0.1200 \\ 
$\chi^2/N_{dof}$  & 5.73/13 & 9.07/16 & 10.8/9      & 3.89/6      & 22.6/14 & 14.5/13  \\  
          & = 0.44 & = 0.57 & = 1.20 & = 0.65 & = 1.61 &  = 1.11 \\
stat.error & 0.0003     & 0.0004      & 0.0003      & 0.0003      & 0.0002 & 0.0003\\ \hline
NLO       & 0.1446     & 0.1437      & 0.1353      & 0.1323      & 0.1454 & 0.1246 \\
$\chi^2/N_{dof}$   & 1.72 & 1.4 & 4.3 & 1.7 & 11.9 & 2.1\\  \hline 
NLO+NLLA  & 0.1271     & 0.1235      & 0.1198      & 0.1164      & 0.1269 & 0.1180 \\
$\chi^2/N_{dof}$   & 0.62 & 1.66 & 7.6 & 7.61 & 2.2 & 5.4 \\  \hline 
%Hybrid    & 0.1280     & 0.1294      & 0.1265      & 0.1243      & 0.1269 & 0.1190 \\ \hline
%ZFITTER   & 0.1236     & 0.1248      & 0.1216      & 0.1195      & 0.1229 & 0.1143\\ \hline
%$x_\mu=0.5$ & 0.1233   & 0.1252      & 0.1239      & 0.1219      & 0.1220 & 0.1169 \\ \hline
%$x_\mu=2.0$ & 0.1324   & 0.1336      & 0.1294      & 0.1270      & 0.1318 & 0.1238\\ \hline
\end{tabular}
\end{center}
\end{table}

%%%%%%%%%%%%%%%%%%%%%%%%%%%%%%%%%%%%%%%%%%%%%%%%%
%%%%% systematic uncertainties
%%%%%%%%%%%%%%%%%%%%%%%%%%%%%%%%%%%%%%%%%%%%%%%%%%

\section{Systematic Uncertainties of \boldmath $\alpha_s$ \unboldmath}
\label{sec:syst}

%%%%%%%%%%%%%%%%%%%%%%%%%%%%%%%%%%%%%%%%%%%%%%%%%
%%%%% exp uncertainties
%%%%%%%%%%%%%%%%%%%%%%%%%%%%%%%%%%%%%%%%%%%%%%%%%%

\subsection{Experimental Uncertainties}
A detailed description of the experimental systematic uncertainties
on the measured event-shape distributions
is found in Ref.~\cite{ALEPH-qcdpaper}. The resulting uncertainties
on the fitted \as{}\ values are estimated
in a  way similar to that for the event shapes themselves. 
Changes of the distributions
under variations of event and particle selection cuts lead in general to small changes in
\as{}. In the fit procedure the same expected statistical
uncertainties (cf.\ section \ref{sec:fits}) are assumed for all variants of the
distribution.
This procedure reduces purely statistical components in the systematic effect, which
are potentially large at LEP2 energies. 
The total experimental systematic uncertainties of \as{}\ 
at LEP2 are between 0.5$\%$ and 1.5$\%$, dominated 
by limitations of the 
correction scheme for initial state radiation. Those at
LEP1 are below 1$\%$ and dominated by imperfections of the
simulation of reconstructed neutral hadronic energy deposits.
As expected, the experimental uncertainties determined when fitting the NNLO 
prediction are generally the same as reported by ALEPH using 
NLO+NLLA fits~\cite{ALEPH-qcdpaper}. 

%%%%%%%%%%%%%%%%%%%%%%%%%%%%%%%%%%%%%%%%%%%%%%%%%
%%%%% theoretical uncertainties
%%%%%%%%%%%%%%%%%%%%%%%%%%%%%%%%%%%%%%%%%%%%%%%%%%

\subsection{Theoretical Uncertainties}
\label{subsec:ther}
\begin{figure}[ht]
\begin{tabular}{lr}
\hspace*{-0.5cm}\includegraphics[width=8cm]{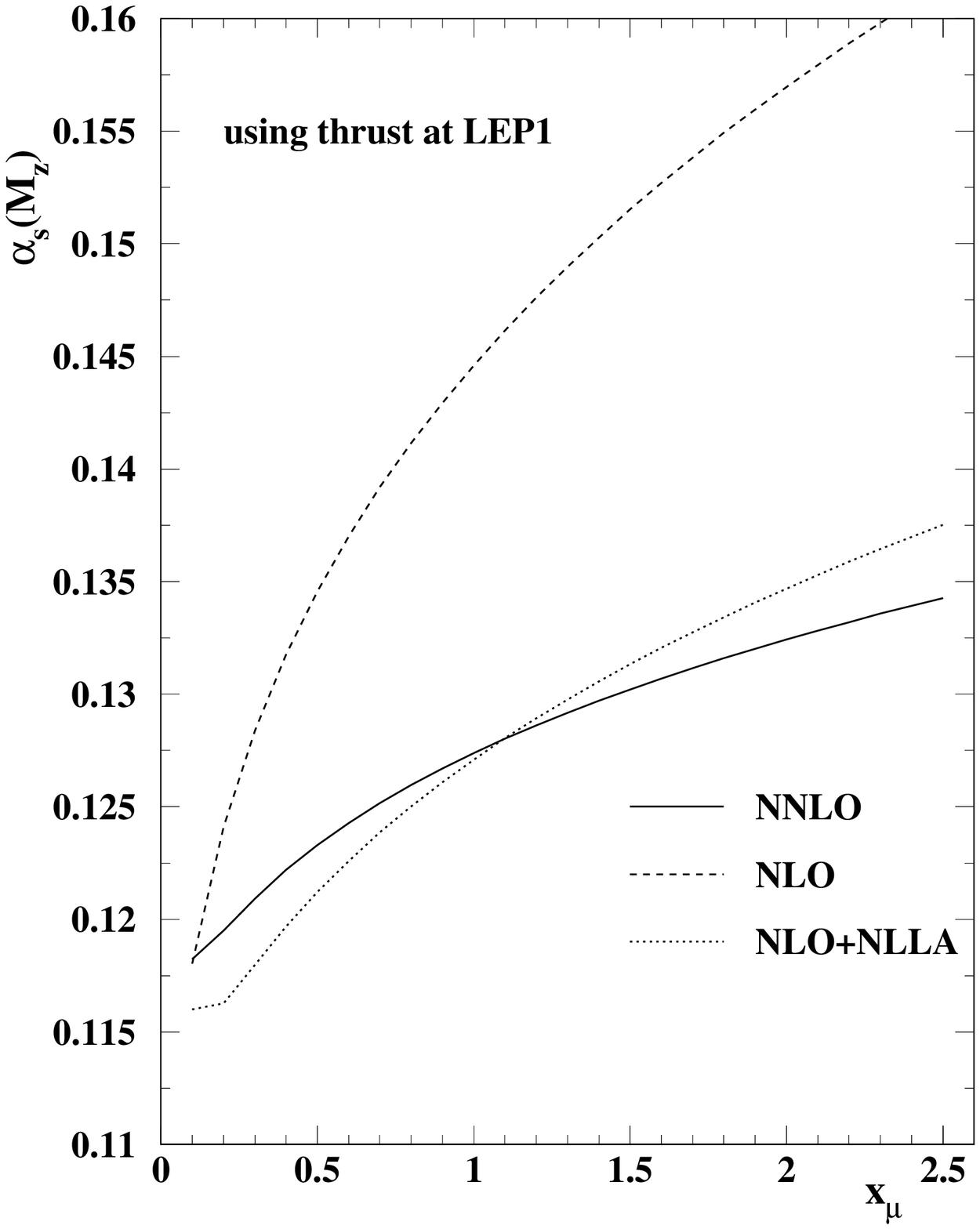} &
\hspace*{-0.5cm}\includegraphics[width=8cm]{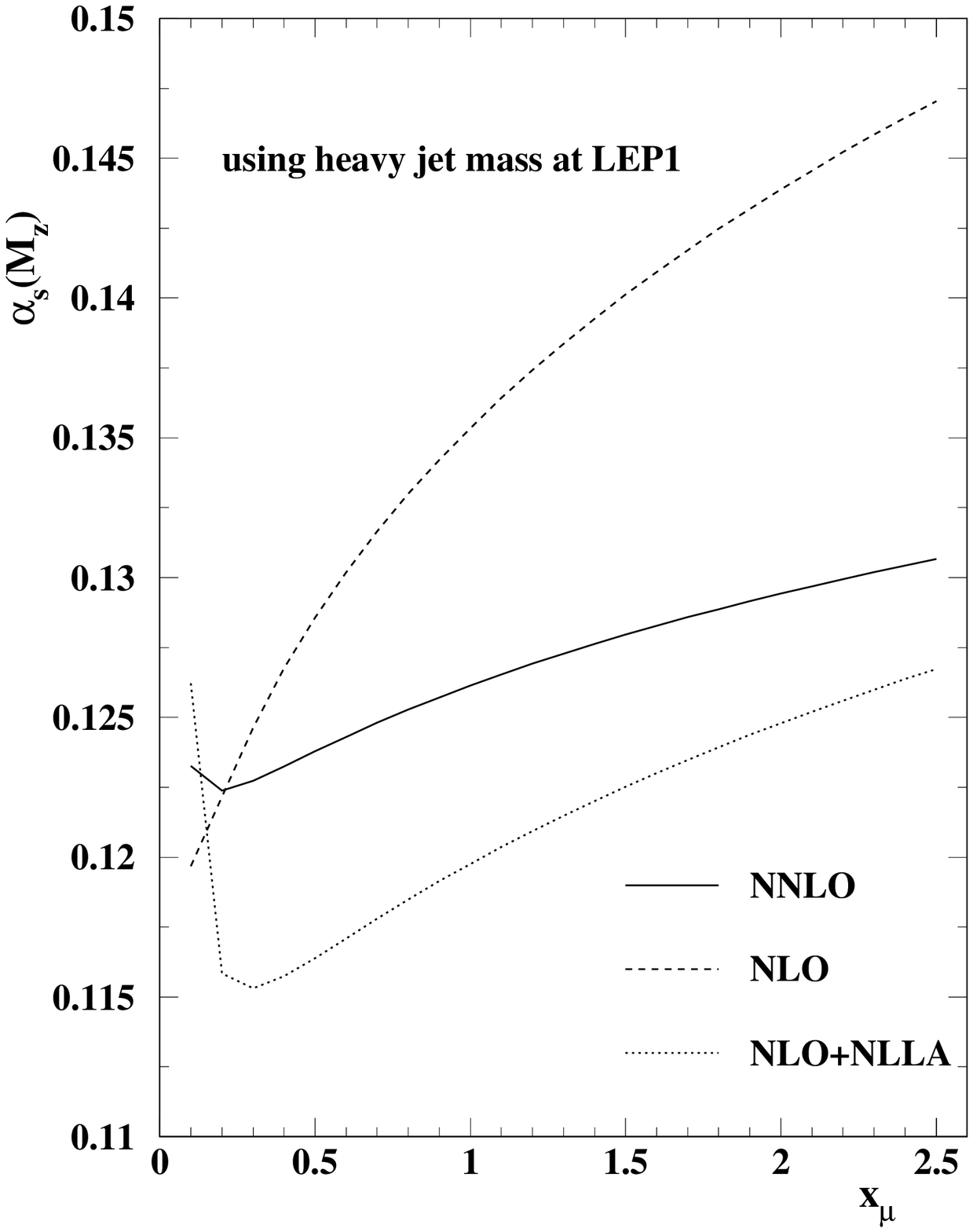} 
\end{tabular}
\caption{\small Dependence of the extracted \as{}\ on the renormalisation scale when fitting 
the thrust (left) and heavy jet mass (right) distributions with predictions at different orders
of perturbation theory.}
\protect\label{fig:scale-var}
\end{figure}

In the case of 
pure fixed-order predictions, the main source
of arbitrariness in the predictions is the choice of the
renormalisation scale $x_\mu$. The residual dependence of the fitted value 
of \as{}(\Mz$^2$) on the renormalisation scale is shown in Fig.~\ref{fig:scale-var}. 
A dramatic reduction of the scale dependence by a factor of two is observed 
when going from NLO to NNLO. But also compared to NLO+NLLA a net improvement 
is obtained when fitting the NNLO prediction.
  
\begin{figure}[ht]
\begin{tabular}{lr}
\hspace*{-0.5cm}\includegraphics[width=8cm]{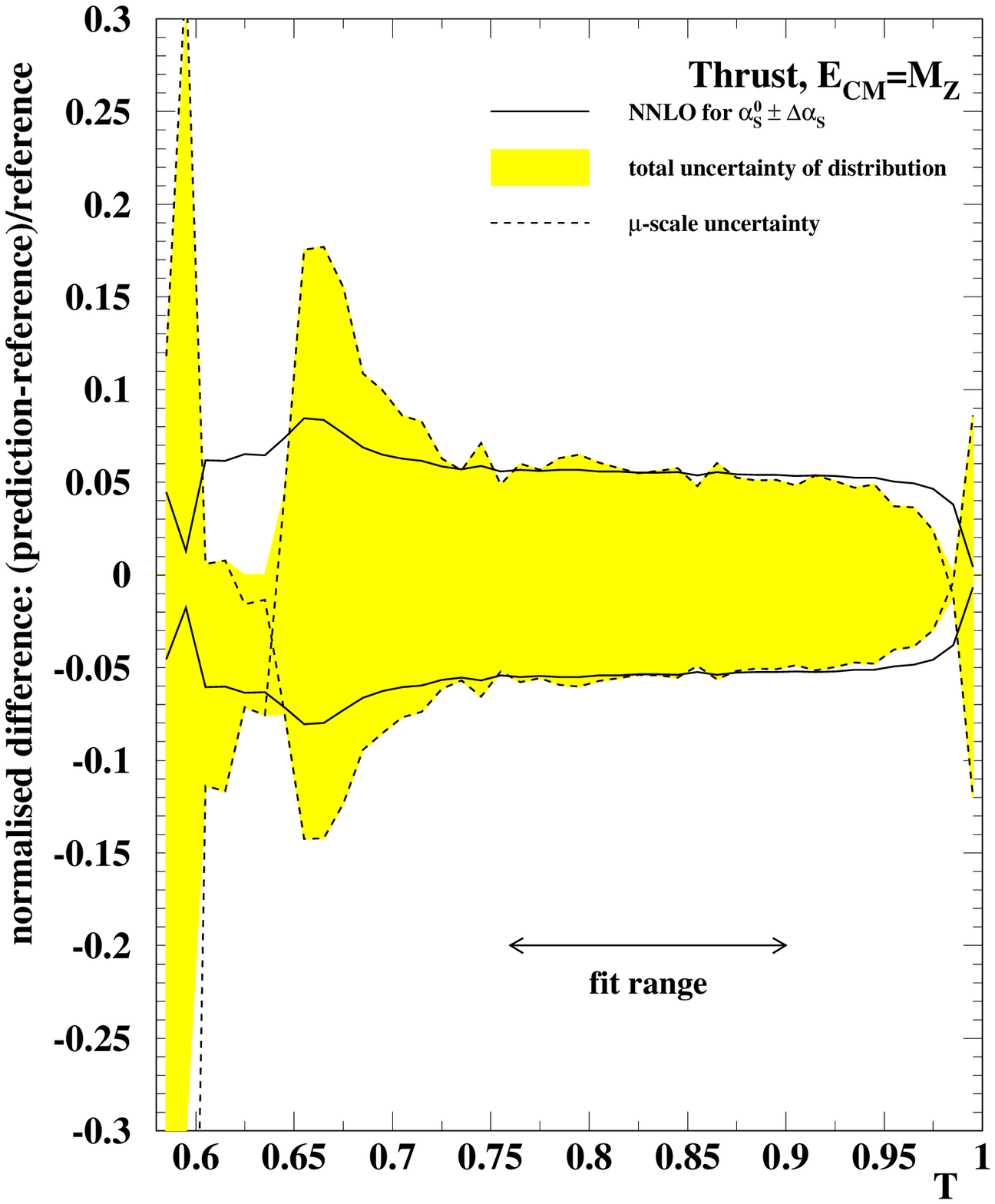} &
\hspace*{-0.5cm}\includegraphics[width=8cm]{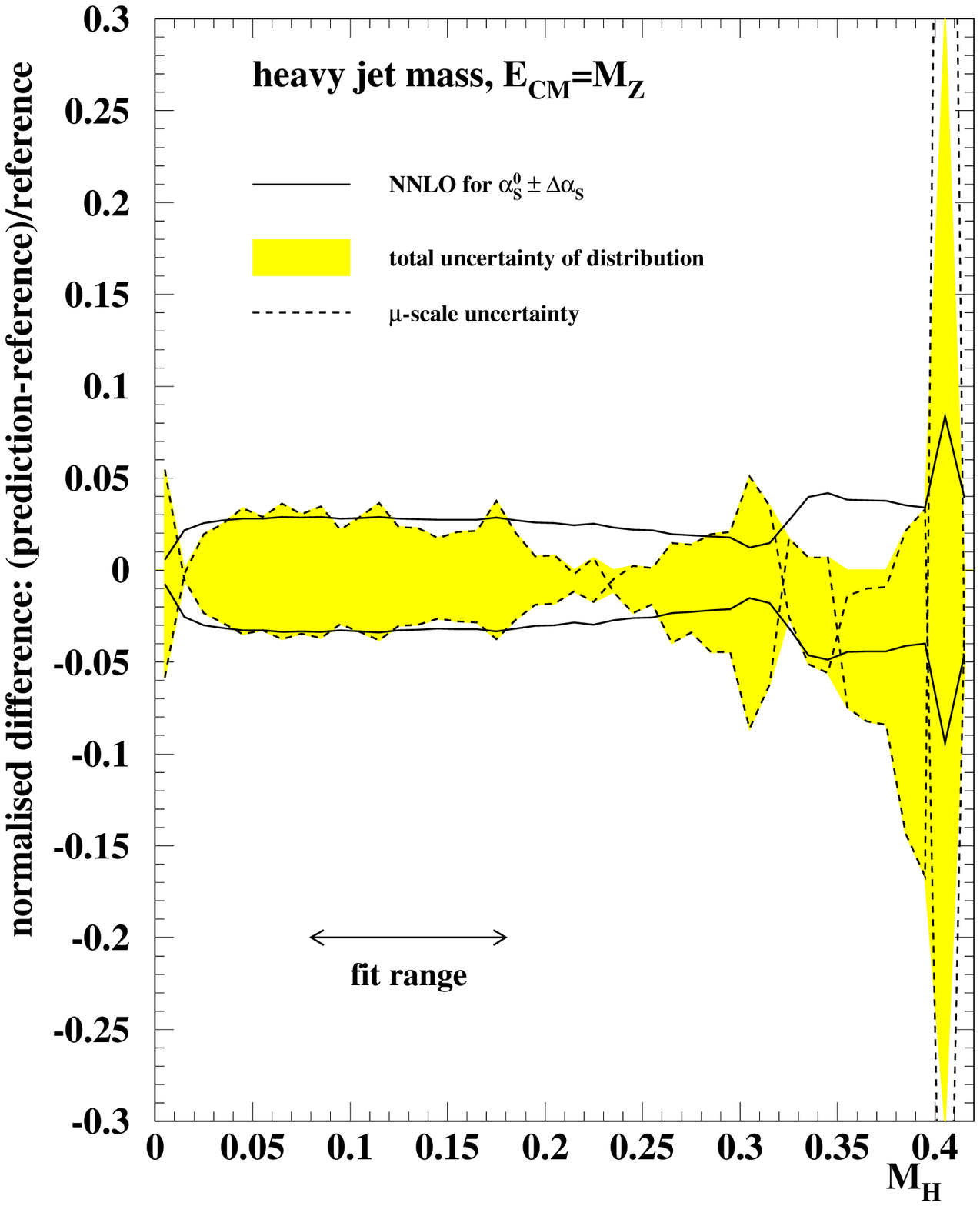} 
\end{tabular}
\caption{\small Theoretical uncertainties for the distributions of
thrust (left) and heavy jet mass (right) at LEP1. The filled area represents the perturbative
uncertainties of the distribution for a given value \as{0}.
The curves show the reference prediction with
$\alpha_s^0 \pm \Delta\alpha_s$. The theoretical uncertainty
$\Delta\alpha_s$ is derived from a fit of the reference theory 
to the contour of the uncertainty band for the actual fit range.}
\protect\label{fig:ththrust}
\end{figure}

The systematic uncertainty related to missing higher orders is estimated with  
the uncertainty-band method recommended in Ref.~\cite{as_theory-uncertainties}.
Briefly, this method derives the uncertainty of \as{}\ from the
uncertainty of the theoretical prediction for the event-shape
distribution and proceeds in three steps. First a reference
perturbative prediction, here NNLO with $x_\mu = 1$, is
determined using the 
value of \as{}\ obtained from the combination of the six 
variables and eight energies, 
as explained in section~\ref{sec:comb}. Then variants of the
prediction with different choices for the $x_\mu$ scale are calculated with the same value of
 \as{}. In each bin of the distribution for a given variable, 
the largest upward
and downward differences with respect to the
reference prediction are taken to define an uncertainty band around the reference theory.
In the last step, the value of \as{}\ in the reference prediction is varied, 
in order to find the range of
values which result in predictions lying inside the uncertainty band
for the fit range under consideration. In contrast to the original 
method \cite{as_theory-uncertainties} we do not request the reference 
prediction to lie strictly inside the uncertainty band, since for the
present NNLO calculations the latter 
is still subject to statistical 
fluctuations. Instead, we make a fit of the reference theory with \as{} as free 
parameter to the uncertainty band, which includes the statistical 
uncertainty on the $C$ coefficient. 
The values of \as{}\ fitted to the upper and lower 
contour of the uncertainty band  
finally set the perturbative systematic uncertainty. The upward and downward
uncertainties are very similar in magnitude and the larger is quoted as symmetric
uncertainty.
The method is
illustrated in Fig. \ref{fig:ththrust}, taking thrust and heavy jet mass as 
examples.

The theoretical error depends on the fit range and on the absolute value of \as{},
scaling approximately with \as{3}\ at NLO and \as{4}\ at NNLO. 
This behaviour can be seen from (\ref{eq:NNLOmu}), where 
variations of $x_\mu$ affect all terms, and individual terms compensate each other across the 
different orders. All compensating terms through $\alpha_s^3$, but no compensating terms 
at $\alpha_s^4$ are included at NNLO, such that any residual dependence 
on $x_\mu$ is to be attributed to these terms.  

At LEP2 energies the statistical fluctuations are large. In order to
avoid biases from downward fluctuations, the theoretical
uncertainties are calculated with the value of \as{}\ 
obtained by the global combination procedure. 
For each energy 
point and in each variable, the error 
is evolved to the 
appropriate energy scale and the uncertainty is 
calculated for the fit range used for the different variables. 

The combined value of \as{}, used to derive the systematic 
perturbative error, depends itself on the theoretical error. 
Hence the procedure of calculating the \as{} combination and 
perturbative error is iterated until convergence is reached, 
typically after two iterations.   

An additional error is evaluated for the b-quark mass correction
procedure. This correction has only been calculated to
$\cal{O}$$(\as{2})$; no resummed and NNLO expressions are yet
available. The difference in \as{}\ obtained with and without
mass corrections is taken as systematic error. The total
perturbative uncertainty quoted in the tables is the quadratic sum
of the errors for missing higher orders and for the mass
correction procedure.
The total perturbative error is between 3\% and 4\% at \Mz\  and 
decreases to between 2\% and 3\% at LEP2 energies. 

The hadronisation model uncertainty is estimated by comparing 
the standard hadron-level event generator programs \HW\
and \AR\ to \PY\ for both hadronisation and detector
corrections. The same set of corrections as in 
Ref.~\cite{ALEPH-qcdpaper} is used.  
Both corrections are calculated with the same generator in order 
to obtain a coherent description at the hadron level.
The maximum change with respect to the nominal result
using \PY\ is taken as systematic error. It was verified that 
calculating the detector corrections
and the hadronization corrections with different event generator programs 
yields consistent results. 
At LEP2 energies the hadronisation
model uncertainty is again subject to statistical fluctuations.
These fluctuations are observed from one energy to the next and
originate from limited statistics of the fully simulated detector-correction functions. Since
non-perturbative effects are expected to decrease with $1/Q$, the
energy evolution of hadronisation errors has been fitted to a simple
$A + B/Q$ parametrisation. The fit was performed for each variable
separately. In the fit procedure a weight scaling with luminosity
is assigned to the hadronisation uncertainty at each energy point.
This ensures that the hadronisation uncertainty at \Mz,
which is basically free of statistical fluctuations, is not
altered by the procedure. As in the case of experimental systematic 
uncertainties, the hadronisation uncertainty is essentially
identical to that published in \cite{ALEPH-qcdpaper}.

The perturbative component of the
error, which is the dominant source of uncertainty in most cases, is highly correlated between the energy
points. The perturbative errors decrease with increasing $Q$, and faster than
the coupling constant itself. The overall error is in general dominated by the 
renormalisation scale dependence.

%%%%%%%%%%%%%%%%%%%%%%%%%%%%%%%%%%%%%%%%%%%%%%%%%
%%%%% combined
%%%%%%%%%%%%%%%%%%%%%%%%%%%%%%%%%%%%%%%%%%%%%%%%%%

\section{Combined Results}
\label{sec:comb}

The measurements obtained with the six different variables are combined into a single
measurement per energy using weighted averages.
The same procedure as in   \cite{ALEPH-qcdpaper}
is applied here and it was verified that the combined results using NLO+NLLA 
with the fit range and weights reported in 
 \cite{ALEPH-qcdpaper} can be reproduced.
A weight is assigned to each observable-dependent measurement $\alpha_s^i$ proportional
to the inverse square of its total error, $ w_i \propto 1/\sigma_i^2$.
The weighted average $\overline{\alpha}_s$ is then given by
\begin{displaymath}
%\label{eq:wgtav}
{\overline \alpha_s}   =   \sum_{i=1}^{N=6}w_i\alpha_s^i\; \;  ,
\end{displaymath}
and the combined statistical error is
\begin{displaymath}
\sigma^{\rm stat}_{{\overline\alpha}_s}  = \sqrt{\sum_{i\neq
j}^{N=6}(\sigma_i w_i)^2 + 2\rho_{ij}\sigma_i w_i\sigma_j w_j } \;
\; ,
\end{displaymath}
for which the correlation coefficients $\rho_{ij}$ are needed.
 This
correlation between fits of $\alpha_s$ to different variables is
obtained using a large number of simulated data samples and
turns out to be typically 60$\%$--80$\%$ (cf.\ Ref.~\cite{ALEPH-qcdpaper}). 
The correlation of systematic errors is taken into account by recomputing the
weighted average for all variations of the analysis, and the
change in $\alpha_s$ with respect to the nominal value is taken as
error.

\begin{table}[hbt]
\caption[Individual results] {\label{tab:Combi}{\small Combined results for
$\alpha_s(Q)$ using NNLO predictions.}}
\begin{center}
\begin{tabular}{|l|cccccccc|}\hline
$Q [$GeV$]$ & 91.2 & 133 & 161 & 172 & 183 & 189 & 200 & 206 \\ \hline 
$\alpha_s(Q)$  & 0.1252 & 0.1202 & 0.1231 & 0.1099 & 0.1122 & 0.1110 & 0.1082 & 0.1064 \\
stat. error    & 0.0002 & 0.0023 & 0.0036 & 0.0048 & 0.0021 & 0.0013 & 0.0014 & 0.0013 \\
exp. error     & 0.0008 & 0.0009 & 0.0009 & 0.0009 & 0.0009 & 0.0009 & 0.0009 & 0.0009 \\
pert. error    & 0.0037 & 0.0030 & 0.0028 & 0.0027 & 0.0026 & 0.0024 & 0.0024 & 0.0023 \\
hadr. error    & 0.0018 & 0.0012 & 0.0009 & 0.0009 & 0.0008 & 0.0008 & 0.0007 & 0.0007 \\
total error    & 0.0042 & 0.0041 & 0.0047 & 0.0057 & 0.0035 & 0.0030 & 0.0030 & 0.0029 \\ \hline 
RMS            & 0.0031 & 0.0014 & 0.0043 & 0.0017 & 0.0030 & 0.0023 & 0.0013 & 0.0023 \\ \hline 
\end{tabular}
\end{center}
\end{table}

\begin{figure}[t!]
\begin{center}
\includegraphics[width=10cm]{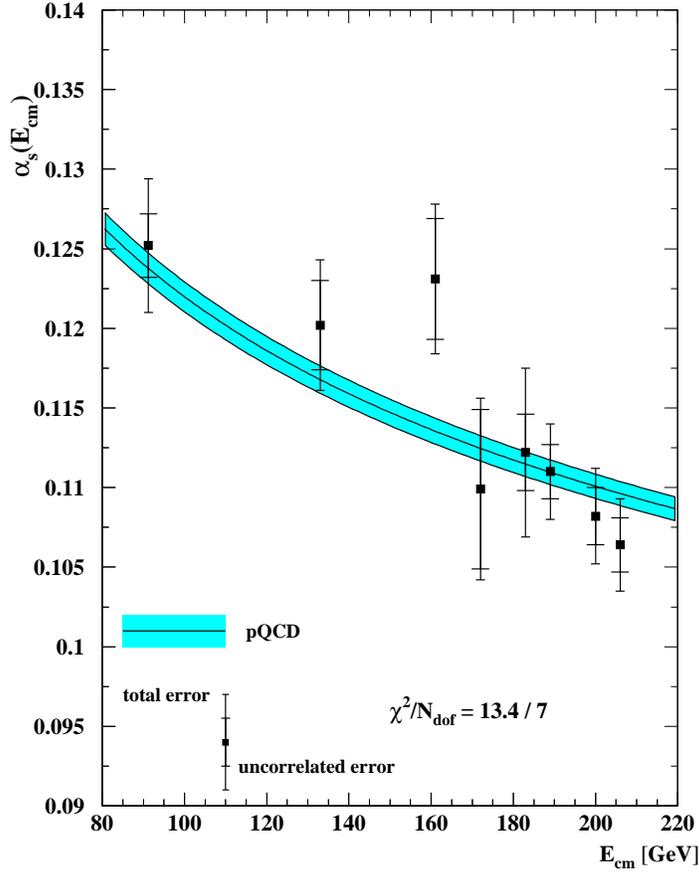}
\end{center}
\caption{\small The measurements of the strong coupling constant $\alpha_s$
between 91.2 and 206 GeV. The results using the six different
event-shape variables are combined with correlations taken into
account. The inner error bars exclude the perturbative uncertainty, which is expected
to be highly correlated between the measurements.
The outer error bars indicate the total error. A fit of the
three-loop evolution formula using the uncorrelated errors is shown. The shaded area
corresponds to the uncertainty in the fit parameter 
$\Lambda^{(5)}_{\overline
{\rm MS}}=284 \pm 14 $ MeV
of the three-loop formula, eq.\ (\protect{\ref{eq:runningas}}).}\protect\label{fig:comb_run}
\end{figure}

The combination of experimental
systematic uncertainties at LEP2 energies is obtained using a
luminosity-weighted average of the uncertainties between 133 GeV
and 206 GeV. Combined results
are given in Table~\ref{tab:Combi}, where 
RMS denotes the root mean squared 
between the different observables. They are  
 shown in Fig.~\ref{fig:comb_run}, together
with a fit of the QCD expectation. The curve is seen to be in good agreement
with the measurements. In the
definition of the $\chi^2$ of the fit only the uncorrelated component of the
errors is taken into account, which excludes the perturbative error.

The combined measurements between 133 and 206 GeV are evaluated at the
scale of the Z boson mass by using the predicted energy
evolution of the coupling constant, eq.\ (\ref{eq:runningas}). The measurements evolved to
\Mz\ are given in Table \ref{tab:combz}. They are again combined using a weighted average, with weights proportional
 to the inverse square of the total uncertainties. In contrast to the combination from different variables, here
the measurements are statistically uncorrelated. Correlations between systematic uncertainties
are taken into account and all variations of the
determination of \as{}\ have been performed for the weighted average.

\begin{table}[hbt]
\caption[Individual results] {\label{tab:combz}{\small Combined results for
$\alpha_s(\Mz)$ using NNLO predictions.}}
\begin{center}
\begin{tabular}{|l|cccccccc|}\hline
$Q [$GeV$]$ & 91.2 & 133 & 161 & 172 & 183 & 189 & 200 & 206 \\ \hline 
$\alpha_s(M_Z)$ & 0.1252 & 0.1276 & 0.1352 & 0.1207 & 0.1247 & 0.1238 & 0.1215 & 0.1196 \\
stat. error     & 0.0002 & 0.0026 & 0.0043 & 0.0058 & 0.0026 & 0.0016 & 0.0017 & 0.0017 \\
exp. error      & 0.0008 & 0.0011 & 0.0012 & 0.0012 & 0.0012 & 0.0011 & 0.0011 & 0.0011 \\
pert. error     & 0.0037 & 0.0032 & 0.0030 & 0.0030 & 0.0028 & 0.0027 & 0.0027 & 0.0027 \\
hadr. error     & 0.0018 & 0.0013 & 0.0011 & 0.0011 & 0.0010 & 0.0009 & 0.0009 & 0.0009 \\
total error     & 0.0042 & 0.0045 & 0.0055 & 0.0067 & 0.0041 & 0.0035 & 0.0035 & 0.0035 \\ \hline 
RMS             & 0.0031 & 0.0015 & 0.0053 & 0.0020 & 0.0037 & 0.0029 & 0.0017 & 0.0029 \\ \hline 
\end{tabular}
\end{center}
\end{table}

The final result is  $\alpha_s(\Mz^2)=0.1240\pm 0.0033$,
and the error components are  given in Table~\ref{tab:comblep}.
Included in Table~\ref{tab:comblep} is
the combination of measurements at LEP2 energies without the point at
\Mz. The total uncertainty of the
combined LEP2 measurements is smaller than the uncertainty 
of the LEP1 measurement,
because the dominant perturbative uncertainties are reduced at higher
energies, even after evolution to \Mz.

\begin{table}[hbt]
\caption[Individual results] {\label{tab:comblep}{\small Weighted average of 
combined measurements for $\alpha_s(\Mz)$ obtained at energies from 91.2 GeV 
to 206 GeV and the average without the point at $\sqrt{s} = \Mz$.}}
\begin{center}
\begin{tabular}{|l|cc|}\hline
data set        & LEP1 + LEP2  & LEP2 \\ \hline
$\alpha_s(M_Z)$ & 0.1240       & 0.1238 \\
stat. error     & 0.0008       & 0.0009 \\
exp. error      & 0.0010       & 0.0011 \\
pert. error     & 0.0029       & 0.0028 \\
hadr. error     & 0.0011       & 0.0010 \\
total error     & 0.0033       & 0.0033 \\ \hline 
\end{tabular}
\end{center}
\end{table}

%%%%%%%%%%%%%%%%%%%%%%%%%%%%%%%%%%%%%%%%%%%%%%%%%%%
%%%%%%% Discussion and Outlook 
%%%%%%%%%%%%%%%%%%%%%%%%%%%%%%%%%%%%%%%%%%%%%%%%%%%

\section{Discussion}
\label{sec:discuss}
The improvements achieved by the NNLO fit compared to that at NLO 
are twofold.
First, the renormalisation scale uncertainty on the extracted value of $\as{}$
is significantly reduced.
This is clearly illustrated in Fig.~\ref{fig:scale-var}
by comparing the NNLO (solid) and NLO (dashed) curves, in the case of
thrust and
heavy jet mass.
Similar improvements are obtained for the other variables.
Second, we find a better description of the  
event-shape distributions over a larger range.  
This is evident in Figs.~\ref{fig:comp_nlo_nnlo1}
and \ref{fig:comp_nlo_nnlo2}.

Of course, we also have to perform a careful comparison of 
the NNLO results with those found in NLO+NLLA analyses, 
which were the state of the art during the LEP era. 
Once again, Fig.~\ref{fig:scale-var} shows that the NNLO
perturbative uncertainty is reduced by about 30\% compared to
NLO+NLLA. 

\begin{figure}[t]
\begin{center}
\includegraphics[width=14cm]{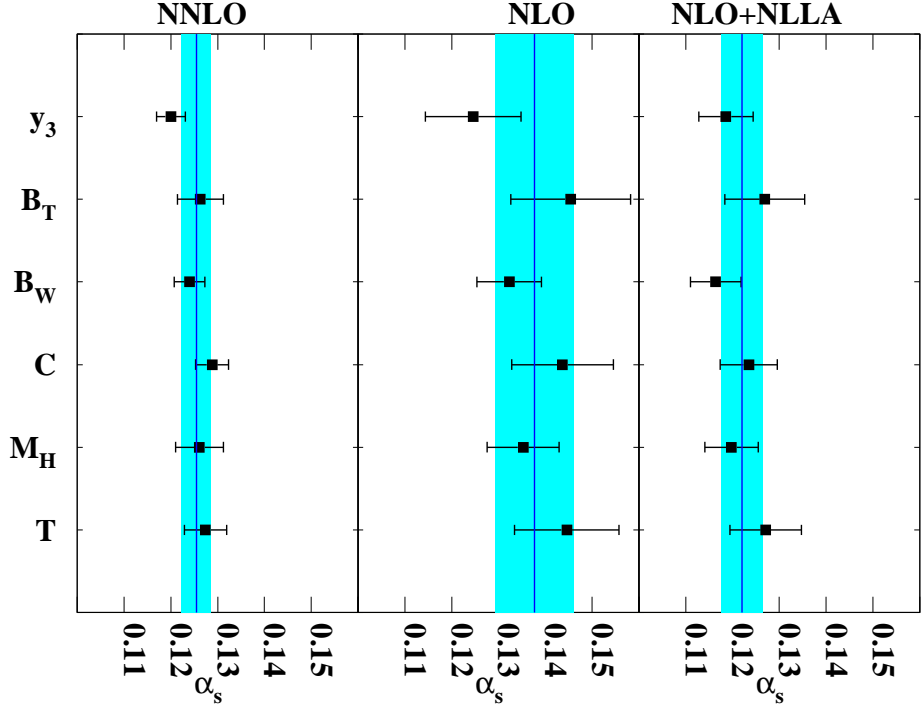}
\end{center}
\caption{\small The measurements of the strong coupling constant $\alpha_s$
for the six event shapes, at $\sqrt{s}=\Mz$, when using QCD predictions at different
approximations in perturbation theory.} \protect\label{fig:scatter}
\end{figure}

It is also remarkable that the \as{}\ values obtained from fits to
different event shapes with
NNLO predictions are considerably more self-consistent than those
found with either
NLO or NLO+NLLA expansions.  
Not only are the extracted values of $\as{}$ more precise, but the spread
obtained from the different observables is smaller.   
This is clearly shown for the data set at $\sqrt{s}=\Mz$ in Fig.~\ref{fig:scatter}.    
The key to this dramatic improvement is the rather different 
size of the NNLO corrections to the
various observables.

Despite these improvements our final combined result on $\alpha_s(M_Z^2)$ still appears 
to be larger than the world average~\cite{as-worldaverage}.
We recall that the value of $\alpha_s(M_Z^2)$ obtained from fits with
NLO+NLLA predictions is smaller than
that obtained with pure NLO calculations alone.
Here we observe that when going from NLO to NNLO there is also a trend in  
the direction of lower values of $\alpha_s(M_Z^2)$.

Clearly, resummed predictions are mandatory in the two-jet region.
Figures~\ref{fig:comp_nlo_nnlo1} and \ref{fig:comp_nlo_nnlo2} clearly show the 
improvement achieved with NLO+NLLA predictions in the 
two-jet region. Measurements 
of \as{} using NLO+NLLA approximations profit from an extended fit range in this region. While a
consistent matching of NNLLA predictions to NNLO would require 
the analytic resummation of next-to-next-to-leading logarithmic terms, 
which are not known at present, a matching of existing 
NLLA expressions to the NNLO calculations requires only the calculation of certain matching 
coefficients~\cite{gionata}. In this case also a better description in the 
two-jet region can be expected.
First preliminary results have been obtained by us with such matched
NNLO+NNLA predictions and fits to 
data seem to confirm the expected trend towards lower values 
of $\alpha_s(M_Z^2)$. We will address this issue in a forthcoming publication.

Electroweak corrections may also be of a similar size as the NNLO corrections
discussed here. At present, these corrections are only known for the
inclusive observables like the hadronic cross section \cite{ZFITTER}, 
where they are found to be sizeable~\cite{hasko_EWQCD}, 
but not for the event-shape distributions. 
This issue deserves further study and will also be addressed in a forthcoming publication.

%%%%%%%%%%%%%%%%%%%%%%%%%%%%%%%%%%%%%%%%%%%%%%%%%
%%%%% Conclusions
%%%%%%%%%%%%%%%%%%%%%%%%%%%%%%%%%%%%%%%%%%%%%%%%%%

\section{Conclusions}
\label{sec:TheEnd}

In this paper we used the newly derived NNLO QCD corrections to event 
shapes in e$^+$e$^-$ annihilation~\cite{ourt,ourevent} to perform 
the first determination of the strong coupling constant $\alpha_s$ 
from event-shape data at NNLO. Our analysis is based on the 
full set of event-shape distributions measured by the ALEPH 
collaboration~\cite{ALEPH-qcdpaper}
at LEP1 and LEP2. 

We observe 
that the inclusion of NNLO QCD corrections to the 
different shape variables yields several important effects, when compared 
to the previously available determinations of $\alpha_s(M_{{\rm Z}}^2)$, 
based either on
pure NLO calculations or NLO predictions matched to NLL approximations~: 
\begin{itemize}
\item[{(a)}] The dominant theoretical uncertainty on  
$\alpha_s(M_{{\rm Z}}^2)$, as estimated from 
scale variations, is reduced by a factor 2 (1.3) compared to 
NLO (NLO+NLLA). A further improvement can be anticipated from a 
matching of NNLO and NLLA predictions.   
\item[{(b)}] The central value obtained at NNLO, 
\begin{center} 
    $\asmz = 0.1240 \;\pm\; 0.0008\,\mathrm{(stat)}
     					 \;\pm\; 0.0010\,\mathrm{(exp)}
                                   \;\pm\; 0.0011\,\mathrm{(had)}
                                   \;\pm\; 0.0029\,\mathrm{(theo)} $,
\end{center}
is about 10\% lower 
than at NLO, and therefore closer to, albeit still larger than, the world average 
from other observables. 
It is also larger than the central value obtained at 
NLO+NLLA~\cite{ALEPH-qcdpaper}, which 
shows the obvious need for a matching of NNLO+NLLA for a fully reliable 
result.  
\item[{(c)}] The scatter among the values of 
$\alpha_s(M_{{\rm Z}}^2)$ extracted 
from the six different event-shape variables is reduced very substantially 
at NNLO. This is a clear indication that this scatter was largely 
due to missing higher order perturbative corrections in previous 
studies. 
\end{itemize}

These observations visibly illustrate the improvements gained from 
the inclusion of the NNLO corrections, and highlight the need for 
further studies on the matching of NNLO+NLLA, on the 
derivation of NNLLA resummation terms, and on the  
electroweak corrections  to event shapes in \epem\emem\ annihilation.

%%%%%%%%%%%%%%%%%%%%%%%%%%%%%%%%%%%%%%%%%%%%%%%%%
%%%%% Thank you
%%%%%%%%%%%%%%%%%%%%%%%%%%%%%%%%%%%%%%%%%%%%%%%%%%

\section*{Acknowledgements}

This research was supported in part by the Swiss National Science Foundation
(SNF) under contract 200020-117602,  
 by the UK Science and Technology Facilities Council and  by the European Commision's Marie-Curie Research Training Network under contract
MRTN-CT-2006-035505 ``Tools and Precision Calculations for Physics Discoveries
at Colliders''.

%%%%%%%%%%%%%%%%%%%%%%%%%%%%%%%%%%%%%%%%%%%%%%%%%
%%%%% bibliography
%%%%%%%%%%%%%%%%%%%%%%%%%%%%%%%%%%%%%%%%%%%%%%%%%%

\bibliographystyle{JHEP}
 
\end{document}